\documentclass[preprint,prd,nofootinbib,tightenlines]{revtex4-1}
\usepackage{graphicx}

\begin{document}
\baselineskip=15pt \parskip=5pt

\vspace*{3em}

\preprint{}

\title{Large Mixing of Light and Heavy Neutrinos in Seesaw Models \\ and the LHC}

\author{Xiao-Gang He, Sechul Oh, Jusak Tandean, Chung-Cheng Wen}
\affiliation{Department of Physics and Center for Theoretical Sciences, \\
National Taiwan University, Taipei 106, Taiwan}

\date{\today $\vphantom{\bigg|_{\bigg|}^|}$}

\begin{abstract}

In the type-I seesaw model the size of mixing between light and heavy neutrinos, $\nu$
and~$N$, respectively, is of order the square root of their mass ratio,
$(m_\nu^{}/m_N^{})^{1/2}$,  with only one generation of the neutrinos.
Since the light-neutrino mass must be less than an~eV or so, the mixing would be very small,
even for a~heavy-neutrino mass of order a~few hundred~GeV.
This would make it unlikely to test the model directly at the LHC, as the amplitude for
producing the heavy neutrino is proportional to the mixing size.
However, it has been realized for some time that, with more than one generation of light
and heavy neutrinos, the mixing can be significantly larger in certain situations.
In this paper we explore this possibility further and consider specific examples in detail
in the context of type-I seesaw.
We study its implications for the single production of the heavy neutrinos at the LHC
via the main channel \,$q\bar q'\to W^*\to\,l N$\, involving an ordinary charged lepton~$l$.
We then extend the discussion to the type-III seesaw model, which has richer phenomenology
due to presence of the charged partners of the heavy neutrinos, and examine
the implications for the single production of these heavy leptons at the LHC.
In the latter model the new kinds of solutions that we find also make it possible to have
sizable flavor-changing neutral-current effects in processes involving ordinary charged leptons.

\end{abstract}


\maketitle

\section{Introduction}

It is now well established from a number of experiments that neutrinos have mass and mix with
each other~\cite{pdg}.  Various ways to go beyond the standard model (SM) have been proposed
in order to accommodate this observation.
Among many possibilities~\cite{zee,seesaw1,seesaw2,Foot:1988aq}, the most popular are
the seesaw scenarios in which new particles are introduced that have masses sufficiently large
to make the light-neutrino masses small.
Needless to say, it is very important to see if models for neutrino masses can be directly
tested experimentally.  The best way to verify the seesaw mechanism directly would be by
observing the heavy particles responsible for generating the tiny neutrino
masses~\cite{colliders,del Aguila:2007em,Franceschini:2008pz,delAguila:2008cj}.
It is widely hoped that this can be realized during the upcoming operation of the CERN Large
Hadron Collider (LHC).  With a center-of-mass energy of 14\,TeV, the $pp$ collisions at the LHC
may lead to the discovery of these heavy particles, thereby providing a window to probe
the models.

In this paper, the seesaw scenarios of interest are the so-called types~I and~III, in which
the heavy particles responsible for the seesaw mechanism are neutral fermions -- the heavy
neutrinos~\cite{seesaw1,Foot:1988aq}.  Whether these heavy neutrinos (as well as their charged
partners in the case of type III) can be produced and detected at colliders crucially depends
on the strength of their interactions with SM particles.  Along this line, we will explore
particularly the possibility of large mixing between the light and heavy neutrinos, subject to
current experimental constraints.   We will first consider the popular type-I seesaw for
a detailed analysis and then extend the discussion to the context of type-III seesaw.

In the type-I seesaw model the size of mixing between the light and heavy neutrinos,
$\nu$ and~$N$, respectively, is of order the square root of their mass ratio,
$(m_\nu^{}/m_N^{})^{1/2}$,  with only one generation of the neutrinos.
Since the light-neutrino mass must be less than an~eV or so, the mixing would be very
small, less than $10^{-5}$ even for $m_N^{}$ of order~100\,GeV.
This would make it impossible to test the model at the LHC, even if it is kinematically
possible for the heavy neutrinos to be singly produced, such as via the quark-level process
\,$q\bar q'\to W^*\to l N$\, involving a SM charged lepton~$l$.
However, it has been realized for some time that, with more than one generation of light and
heavy neutrinos, there are circumstances in which the mixing can be much larger~\cite{large-mix},
offering greater hope to test the seesaw mechanism at the LHC.
In this paper, we explore this possibility further and consider specific examples in detail
in the context of type-I seesaw.
We examine its implications for the single production of $N$ at the LHC via the main channel
\,$q\bar q'\to W^*\to l N$.\,
Subsequently we extend the analysis to the situation in the type-III seesaw model, which has
richer phenomenology due to presence of the charged partners $E$ of the heavy neutrinos.
We then discuss the implications for the single production of these heavy leptons at the LHC,
via \,$q\bar q'\to W^*\to l N$\, and \,$q\bar q\to Z^*\to l E$.\,
Interestingly, in type-III seesaw the new kinds of solutions that we find also make it
possible to have sizable flavor-changing neutral current (FCNC) effects in processes
involving SM charged leptons.

In the type-I seesaw scenario, the seesaw mechanism is realized by introducing
right-handed neutrinos that are singlets under the SM gauge groups and can therefore have
large Majorana masses~\cite{seesaw1}.
For one generation of light and heavy neutrinos, $\nu_L^{}$ and $N_R^{}$, the relevant
Lagrangian describing their masses is given by
\begin{eqnarray}
{\cal L} \,\,=\,\, - \bar N_R^{}\, Y_D^{} \tilde H^\dagger L_L^{}
- \mbox{$\frac{1}{2}$} \bar N_R^{}\, M_N^{} (N_R^{})^c \,\,+\,\, {\rm H.c.} ~,
\end{eqnarray}
where  $Y_D^{}$ is the Yukawa coupling, \,$\tilde H=i\tau_2^{}H^*$\,  with
$\tau_2^{}$ being the usual second Pauli matrix and
\,$H=\bigl(\phi^+~~\phi^0\bigr){}^{\rm T}=\bigl(\phi^+~~(v+h+i\eta)/\sqrt2\bigr){}^{\rm T}$\,
the Higgs doublet having vacuum expectation value~$v$,
\,$L_L^{}=\bigl(\nu_L^{}~~l^-_L\bigr){}^{\rm T}$\, is the left-handed lepton doublet,
$M_N$ is the Majorana mass of $N_R$, and $(N_R)^c$\, denotes the charge conjugate of $N_R$.
The Dirac mass in this case is therefore \,$m_D^{}=v Y_D^{}/\sqrt2$.\,
With more than one generation of light and heavy neutrinos, the resulting mass terms can be
expressed~as
\begin{eqnarray} \label{Lmass}
{\cal L}_{\rm mass}^{} \,\,=\,\, -\mbox{$\frac{1}{2}$}
\Bigl( \overline{\bigl(\nu_L^{}\bigr)^{c}} \hspace{3ex} \bar N_R^{} \Bigr)
M_{\rm seesaw}^{} \left( \begin{array}{c} \nu_L^{} \vspace{1ex} \\
\bigl(N_R^{}\bigr)^c \end{array} \right)
\,\,+\,\, {\rm H.c.} ~,
\end{eqnarray}
with the seesaw mass matrix
\begin{eqnarray}
M_{\rm seesaw}^{} \,\,=\,\,
\left( \begin{array}{cc} 0 & m_D^{\rm T} \vspace{0.5ex} \\ m_D^{} & M_N^{}
\end{array} \right) ,
\end{eqnarray}
where now $\nu_L^{}$ and $N_R^{}$ are column matrices and $m_D^{}$ and $M_N^{}$ are
square matrices, $M_N^{}$ also being symmetric.
Without loss of generality, in what follows we work in the basis where $M_N$ is already
diagonal and real, unless otherwise indicated.

One can write the weak eigenstates $\nu_L^{}$ and $(N_R)^c$ in terms of the mass eigenstates
$\nu_{mL}^{}$ and $N_{mL}^{}$~as
\begin{eqnarray} \label{basis}
\left( \begin{array}{c} \nu_L^{} \vspace{1ex} \\ \bigl(N_R^{}\bigr)^c \end{array} \right)
=\,\, U \left( \begin{array}{c} \nu_{mL}^{} \vspace{1ex} \\ N_{mL}^{} \end{array} \right) ,
\hspace{5ex}
U \,\,\equiv\,\, \left(\begin{array}{cc} U_{\nu\nu}^{} & U_{\nu N}^{} \vspace{1ex} \\
U_{N \nu}^{} & U_{NN}^{} \end{array}\right) .
\end{eqnarray}
Thus $U$ is unitary and diagonalizes $M_{\rm seesaw}$,
\begin{eqnarray} \label{mdiag}
\left(\begin{array}{cc} \hat m_\nu^{} & 0 \vspace{0.5ex} \\ 0 & \hat M_N^{} \end{array}\right)
=\,\, U^{\rm T}  M_{\rm seesaw}^{} U ~,
\end{eqnarray}
where for three generations
\begin{eqnarray}
\hat m_\nu^{} = {\rm diag}\bigl(m_{\nu_1},m_{\nu_2},m_{\nu_3}\bigr) ~, \hspace{5ex}
\hat M_N^{} = {\rm diag}\bigl(M_1, M_2, M_3\bigr) ~.
\end{eqnarray}
On the other hand, the block matrices  $U_{\nu\nu}$, $U_{\nu N}$, $U_{N \nu}$, and $U_{NN}$
are not unitary.
Assuming that the nonzero elements of $M_N^{}$ are all much larger than those of $m_D^{}$,
and expanding in terms of $m_D^{}M_N^{-1}$, one then finds to leading order
\begin{eqnarray} \label{U}
U_{\nu N}^{} \,\,=\,\, m_D^\dagger\, \hat M_N^{-1} ~, \hspace{5ex}
U_{N\nu}^{} \,\,=\,\, -M_N^{-1}\, m_D^{}\, U_{\nu\nu}^{} ~, \hspace{5ex}
U_{NN}^{} \,\,=\,\, 1 ~,
\end{eqnarray}
$U_{\nu\nu}^{}$ having small deviations from unitarity, and the reduced light-neutrino mass
matrix
\begin{eqnarray} \label{mnud}
m_\nu^{} \,\,\equiv\,\, -m_D^\dagger\, \hat M_N^{-1}\, m_D^* ~.
\end{eqnarray}
The matrix $m_\nu^{}$ can be diagonalized using the unitary Pontecorvo-Maki-Nakagawa-Sakata
matrix $U_{\rm PMNS}$~\cite{pmns},
\begin{eqnarray}
\hat m_\nu^{} \,\,=\,\, U_{\rm PMNS}^\dagger\, m_\nu^{}\, U_{\rm PMNS}^* ~.
\end{eqnarray}
Since $U_{\nu\nu}$ is nearly unitary and plays the role of $U_{\rm PMNS}$ in a theory with only
three light neutrinos, for numerical analysis we will take $U_{\nu\nu}$ to be $U_{\rm PMNS}$.

In terms of the weak eigenstates, the neutrinos couple to the gauge and Higgs bosons in
the SM according to
\begin{eqnarray}
{\cal L}' \,\,=\,\,
\biggl( \frac{g}{\sqrt2}\, \bar l_L^{}\, \gamma^\mu \nu_L^{} W^-_\mu
- \bar N_R^{}\, m_D^{} \nu_L^{}\, \frac{h}{v} \,+\, {\rm H.c.} \biggr)
\,+\, \frac{g}{2 c_{\rm w}^{}}\, \bar\nu_L^{} \gamma^\mu \nu_L^{} Z_\mu^{} ~,
\end{eqnarray}
where  \,$c_{\rm w}^{}=\cos\theta_{\rm W}^{}$.\,
In the mass-eigenstate basis, given in Eq.~(\ref{basis}), one can rewrite ${\cal L}'$ as
\begin{eqnarray} \label{L'}
{\cal L}' &=&
\frac{g}{\sqrt2} \Bigl(
\bar l_L^{}\, \gamma^\mu U_{\nu\nu}^{} \nu_{mL}^{} W^-_\mu +
\bar l_L^{}\, \gamma^\mu U_{\nu N}^{} N_{mL} ^{} W^-_\mu \,+\, {\rm H.c.} \Bigr)
\nonumber \\ && +\,\,
\frac{g}{2 c_{\rm w}^{}} \Bigl(
\bar\nu_{mL}^{}\, \gamma^\mu U^\dagger_{\nu\nu} U_{\nu\nu}^{} \nu_{mL}^{}
+ \bar N_{mL}^{}\, \gamma^\mu U_{\nu N}^\dagger U_{\nu\nu}^{} \nu_{mL}^{}
\nonumber \\ && \hspace*{8ex} +\,\,
\bar\nu_{mL}^{}\, \gamma^\mu U^\dagger_{\nu\nu} U_{\nu N}^{} N_{mL}^{}
+ \bar N_{mL}^{}\, \gamma^\mu U_{\nu N}^\dagger U_{\nu N}^{} N_{mL}^{} \Bigr) Z_\mu^{}
\nonumber\\ && -\,\, \Bigl[
\overline{\bigl(\nu_{mL}^{}\bigr)^{\!c}}\,\hat m_\nu^{}U_{\nu\nu}^\dagger U_{\nu\nu}^{}\nu_{mL}^{}
+ \overline{\bigl(N_{mL}^{}\bigr)^{\!c}}\,\hat M_N^{} U_{\nu N}^\dagger U_{\nu\nu}^{} \nu_{mL}^{}
\nonumber \\ && \hspace*{4ex} +\,\,
\overline{\bigl(\nu_{mL}^{}\bigr)^{\!c}}\,\hat m_\nu^{} U_{\nu\nu}^\dagger U_{\nu N}^{} N_{mL}^{}
+ \overline{\bigl(N_{mL}^{}\bigr)^{\!c}}\,\hat M_N^{} U_{\nu N}^\dagger U_{\nu N}^{} N_{mL}^{}
\,+\, {\rm H.c.} \Bigr] \frac{h}{v}
\end{eqnarray}
using the relations \,$U_{N\nu}^{\rm T}\, m_D^{}=\hat m_\nu^{} U_{\nu\nu}^\dagger$\,
and  \,$U_{NN}^{\rm T}\, m_D^{}=\hat M_N^{}\, U_{\nu N}^\dagger$\,  derived from
Eq.~(\ref{mdiag}).

\begin{figure}[ht]
\includegraphics[width=4in]{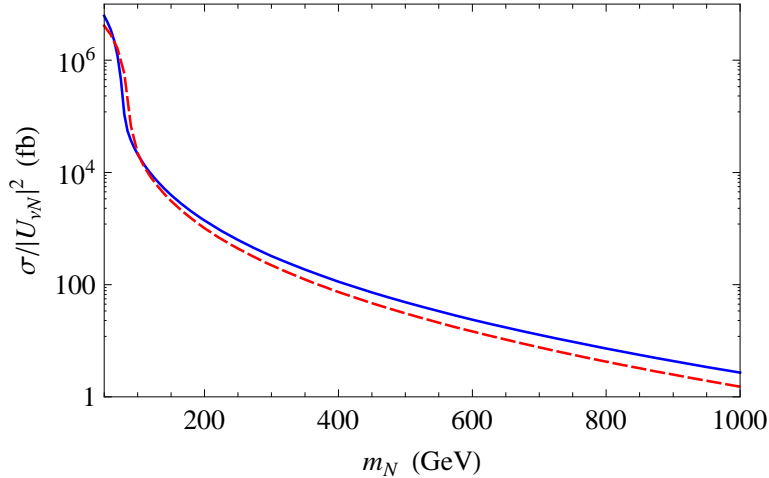}
\caption{Cross sections of \,$p p\to l N X$\, in types-I and -III seesaw (solid curve)
and \,$pp\to l E X$\, in type-III seesaw (dashed curve)\, as functions of \,$m_N^{}=m_E^{}$,\,
for $pp$ center-of-mass energy of \,$\sqrt s=14$\,TeV,\, with
$\sigma/|U_{\nu N}|^2$ indicating that the $U_{\nu N}$ elements associated with the $WlN$ and
$ZlE$ couplings have been set to unity in the cross sections.\label{csplot}}
\end{figure}

The equations for ${\cal L}'$ above indicate that, although the heavy neutrinos $N_R$ do not
directly have SM gauge interactions, through mixing they can interact with the SM gauge bosons.
In particular, $N$~couples to $W$, $Z$, or~$h$ at tree level, as can be seen from Eq.~(\ref{L'}).
Consequently, $N$~can be singly produced via the quark-level processes
\,$q\bar q'\to W^*\to l N$\, and  \,$q\bar q\to\bigl(h^*,Z^*\bigr)\to\nu N$,\,
the former with a charged lepton $l$ in the final state being easier to observe.
This implies that the LHC can, in principle, test the seesaw mechanism for $N$-mass values up
to a TeV or so.
Since all these $N$-production processes depend on the elements of $U_{\nu N}$, their size
plays a crucial role as far as the testability of the seesaw mechanism is concerned.
In Fig.~\ref{csplot}, we show the cross section of \,$p p\to l N X$\, (solid curve) as
a function of $m_N^{}$ for $pp$ center-of-mass energy of \,$\sqrt s=14$\,TeV,\,
with the $U_{\nu N}$ element associated with the $WlN$ coupling set to unity.
It is therefore interesting to examine how large the elements of $U_{\nu N}$ can be, taking
into account constraints from existing experimental data, especially the light-neutrino masses.
This we will do in the next section.

From Eq.~(\ref{L'}), one can also extract the couplings of $N$ to SM particles in order to
evaluate the rates of its decay modes.
We have collected the formulas for the dominant modes in Appendix~\ref{widths}, which each
depend on the elements of  $U_{\nu N}$.
We will illustrate this dependence in the examples studied in the next section.
Thus, once $N$ is discovered, one may gain additional information about neutrino mixing patterns
by studying its individual decays and their branching ratios.

\section{Small light-neutrino masses and large light-heavy mixing}

With only one generation and the requirement \,$m_N^{}\gg|m_D^{}|$\, for the $N$ mass,
the light-neutrino mass is given by \,$m_\nu^{}= - m^2_D/m_N^{}$\, at leading order,
which explains why the light-neutrino mass is much smaller than the mass of its
charged-lepton partner.
For one generation, the mixing between the light and heavy neutrinos has a magnitude of
\,$|U_{\nu N}^{}|=|m_D^{}/m_N^{}|=(m_\nu^{}/m_N^{})^{1/2}$.\,
If $|U_{\nu N}|^2$ is large enough, the heavy neutrino $N$  may be produced at the LHC.
However, with the light-neutrino mass constrained to be less than ${\cal O}$(1\,eV),
the size of $U_{\nu N}$ is bounded by~\,$10^{-5}(100{\rm\,GeV}/m_N^{})^{1/2}$.
Hence the mixing is extremely small, even with $m_N^{}$ as low as~100\,GeV.
With such small mixing, it is not possible to produce enough number of heavy neutrinos
to study its properties at the LHC.
This may lead one to think naively that it is not possible to test type-I seesaw at the LHC.
However, before drawing such a conclusion, one should make sure that with more than one
generation the elements of $U_{\nu N}$ are always constrained to be as small as that with
just one generation.
We find that this is not generally true and that it is possible to have large enough
$U_{\nu N}$ such that testing the seesaw mechanism at the LHC can be achieved.

With more than one generation, from Eqs.~(\ref{U}) and~(\ref{mnud}) we have the leading-order
relation
\begin{eqnarray} \label{UmU}
U_{\rm PMNS}^{}\, \hat m_\nu^{}\, U_{\rm PMNS}^{\rm T} \,\,=\,\,
-U_{\nu N}^{}\, \hat M_N^{}\, U_{\nu N}^{\rm T} ~.
\end{eqnarray}
Thus, if there is a nontrivial solution for $U_{\nu N}$ which makes the right-hand side
vanish exactly, the elements of $U_{\nu N}$ can be large and evade the constrain
\,$|U_{\nu N}|^2 = m_\nu^{}/m_N^{}$\, in the one-generation case~\cite{large-mix}.
We will denote such a solution by $U_0$.  Once $U_0$ is found, one should of course make sure
that by adding a perturbation $U_\delta$ to $U_0$, so that \,$U_{\nu N}=U_0+U_\delta$,\,
one can obtain the light-neutrino masses and mixing.
As we discuss below, such solutions indeed exist.
In the following we will work with three generations of light and heavy neutrinos.

Let us first consider what the rank of $U_0$ must be that can yield the right solutions.
We find it convenient to work in the basis where $U_0$ is already diagonalized.
In that case, $M_N^{}$ generally is not diagonal.
Since $M_N^{}$ must be of rank three in order that the three heavy neutrinos have nonzero
masses, the rank of $U_0^{}$ must not be more than two, for otherwise the determinant of
\,$U_0^{}M_N^{}U_0^{\rm T}$\, would not be zero in contradiction with the assumption
\,$U_0^{}M_N^{}U_0^{\rm T}=0$.\,
It turns out that $U_0^{}$ of rank two is also problematic.
Writing the diagonal form of $U_0$ as \,$\hat U_0={\rm diag}(a,b,0)$,\, we have
\begin{eqnarray}
\hat U_0^{} M_N^{}\hat U_0^{\rm T} \,\,=\, \left( \begin{array}{ccc}
a^2\, M_{11}^{} \,&\, a b\, M_{12}^{} \,&\, 0 \vspace{0.5ex} \\ \vspace{0.5ex}
a b\, M_{12}^{} \,&\, b^2\, M_{22}^{} \,&\, 0 \\ 0 \,&\, 0 \,&\, 0 \end{array} \right) ,
\end{eqnarray}
where $M_{ij}$ are the elements in $M_N$, which is symmetric.
Since $M_N$ is of rank three, $M_{12,11,22}$ cannot all be simultaneously zero.
This implies that, if we keep at least one of $a$ and $b$ nonzero, we have two types of
nontrivial solutions with necessary and sufficient conditions:
(1)~$b=0$,\, $M_{11}^{}=0$\, and \,(2)~$a=0$,\, $M_{22}^{}=0$.\,
We conclude that $U_0$ must be a rank-one matrix.
This type of texture for the mass matrix can be made stable by imposing
symmetries~\cite{large-mix,He:2009xd}.

Next we derive the general expression for $U_0^{}$ which has rank one and satisfies
\begin{eqnarray} \label{zero}
U_0^{}\hat M_N^{}U_0^{\rm T} \,\,=\,\, 0 ~,
\end{eqnarray}
where  \,$\hat M_N^{}={\rm diag}\bigl(M_1^{}, M_2^{}, M_3^{}\bigr)$\, as before.
Without loss of generality, the diagonal form of $U_0^{}$ can be chosen to be
\,$\hat U_0^{}={\rm diag}(\hat u,0,0)$,\,  with $\hat u$ being some constant.
This is related to the nondiagonal $U_0^{}$ by the biunitary transformation
\,$U_0^{}=V'\hat U_0^{}V$,\,  where $V$ and $V'$  are 3$\times$3 unitary matrices.
As a consequence, $U_0^{}$ and $\hat U_0^{}$ share the same rank.
Denoting the elements of $V^{(\prime)}$ by $V_{kl}^{(\prime)}$, we then arrive~at
\begin{eqnarray} \label{U0gen}
U_0^{} \,\,=\,\, \kappa\left( \begin{array}{ccc}
a\, V_{11}^{} \,&\, a\, V_{12}^{} \,&\, a\, V_{13}^{} \vspace{0.5ex} \\ \vspace{0.5ex}
b\, V_{11}^{} \,&\, b\, V_{12}^{} \,&\, b\, V_{13}^{} \\
c\, V_{11}^{} \,&\, c\, V_{12}^{} \,&\, c\, V_{13}^{} \end{array} \right) ,
\end{eqnarray}
where $\kappa$ is a proportionality constant, \,$\kappa a=\hat u V_{11}'$,\,
\,$\kappa b=\hat u V_{21}'$,\, and  \,$\kappa c=\hat u V_{31}'$.\,
Since this $U_0^{}$ has to satisfy Eq.~(\ref{zero}), we use the relation
\,$U_0^{}=V'\hat U_0^{}V$\, in the equation to find \,$\hat U_0^{}M_N^{}\hat U_0^{\rm T}=0$,\,
where \,$M_N^{}=V\hat M_N^{}V^{\rm T}$,\, which is clearly symmetric and generally nondiagonal.
As shown in the preceding paragraph, this implies that the (1,1) element of $M_N^{}$ must
vanish, \,$M_{11}^{}=0$,\, which translates into
\begin{eqnarray} \label{MV2}
M_1^{}\,V_{11}^2 + M_2^{}\,V_{12}^2 + M_3^{}\,V_{13}^2 \,\,=\,\, 0 ~.
\end{eqnarray}
Thus, with $V_{11}$, $V_{12}$, and $V_{13}$ required to fulfill this condition,
Eq.~(\ref{U0gen}) has the desired expression for $U_0$ in terms of the parameters
$a$, $b$, and $c$, which are to be fixed from experimental data.

To illustrate that one can find $U_0^{}$ of the form in Eq.~(\ref{U0gen}) that yields
large light-heavy mixing while simultaneously satisfying constraints from various
measurements, we find it convenient to consider specific examples.
Accordingly, in the rest of this section we work in the basis in which $M_N$ is diagonal,
writing
\begin{eqnarray} \label{Mdiag}
M_N^{} \,\,=\,\, \hat M_N^{} \,\,=\,\,
{\rm diag}\Biggl(\frac{1}{r_1^{}},\,\frac{1}{r_2^{}},\,\frac{1}{r_3^{}}\Biggr)m_N^{} ~,
\hspace{5ex} r_i^{} \,\,=\,\, \frac{m_N^{}}{M_i^{}} ~,
\end{eqnarray}
where we have regarded $m_N^{}$ as representative of the mass scale of the heavy neutrinos,
and so can take it to be the lightest of $M_{1,2,3}$.
Then, choosing the appropriate $V_{11,12,13}$ in Eq.~(\ref{U0gen}) to satisfy Eq.~(\ref{MV2}),
as well as the unitarity relation \,$|V_{11}|^2+|V_{12}|^2+|V_{13}|^2=1$,\,
and adjusting $\kappa$ in each instance, we obtain as examples for $U_0$
\begin{eqnarray} \label{U0}
U_0^a &=& \left( \begin{array}{lll}
a \,&\, a \,&\, i \sqrt2\, a \vspace{0.5ex} \\ b \,&\, b \,&\, i\sqrt2\, b \vspace{0.5ex} \\
c \,&\, c \,&\, i\sqrt2\, c \end{array}\right) \!{\cal R} ~, \hspace{5ex}
U_0^b \,\,=\, \left( \begin{array}{lll} a & i a & 0 \vspace{0.5ex} \\
b & i b & 0 \vspace{0.5ex} \\ c & i c & 0 \end{array}\right) \!{\cal R} ~, \hspace{5ex}
U_0^c \,\,=\, \left( \begin{array}{lll} a & 0 & i a \vspace{0.5ex} \\
b & 0 & i b \vspace{0.5ex} \\ c & 0 & i c \end{array}\right) \!{\cal R} ~,
\nonumber \\
U_0^d &=& \left( \begin{array}{ccc} 0 \,&\, a \,&\, i a \vspace{0.5ex} \\
0 & b & i b \vspace{0.5ex} \\ 0 & c & i c \end{array}\right)^{\vphantom{|}} \!{\cal R} ~,
\hspace{5ex}
{\cal R} \,\,=\,\, {\rm diag}\Bigl(\sqrt{r_1^{}},\,\sqrt{r_2^{}},\,\sqrt{r_3^{}}\Bigr) ~,
\end{eqnarray}
where we have factored out the diagonal matrix $\cal R$ in order to maintain the form of
$U_0$ in cases where the heavy neutrinos are nondegenerate.
In our numerical examples, we will use in particular only  $U_0^a$ and $U_0^d$, although
the other options would also be possible.
We remark that it may sometimes be necessary to use trial and error in order to determine
the right choice of $U_0$, with $a$, $b$, and $c$ being subject to experimental constraints
that are relevant in specific circumstances.

We now show explicitly that one can find solutions for $U_{\nu N}$ which satisfy all
experimental data on light neutrinos by adding a perturbation matrix $U_\delta$ given by
\begin{eqnarray}
U_{\delta}^{} \,\,=\, \left(\begin{array}{lll}
\delta_{11}^{} & \delta_{12}^{} & \delta_{13}^{} \vspace{0.5ex} \\
\delta_{21}^{} & \delta_{22}^{} & \delta_{23}^{} \vspace{0.5ex} \\
\delta_{31}^{} & \delta_{32}^{} & \delta_{33}^{} \end{array} \right) \!{\cal R} ~.
\end{eqnarray}
Since  \,$U_0^{}\hat M_N^{} U_0^{\rm T} = 0$,\, the light-neutrino mass matrix
\,$m_\nu^{}=-U_{\nu N}^{}\, \hat M_N^{}\, U_{\nu N}^{\rm T}$\, becomes
\begin{eqnarray} \label{mnu}
m_\nu^{} \,\,=\,\, -U_0^{}\hat M_N^{}U^{\rm T}_\delta
- U_\delta^{}\hat M_N^{}U_0^{\rm T} - U_\delta^{}\hat M_N^{}U_\delta^{\rm T} ~.
\end{eqnarray}
When trying to find solutions, one should keep in mind that $\delta_{ij}$ as perturbations
should be much smaller than $a$, $b$, and $c$, while the size of  \,$(a,b,c)\delta_{kl}M_i$\,
is set by the light-neutrino masses.  One can neglect the last term,
$U_\delta^{} M_N^{} U^{\rm T}_\delta$, in Eq.~(\ref{mnu}) compared with the first two terms.
In case the combination of the first two terms in Eq.~(\ref{mnu}) happens to vanish,
one needs to keep the $U_\delta^{} M_N^{} U^{\rm T}_\delta$ term, the elements of
$U_\delta$ now being of order $(m_\nu^{}/m_N^{})^{1/2}$.

\subsection{Solutions with one of the light-neutrino masses being zero}

Defining
\begin{eqnarray} \label{tu}
\bar U_{\nu N}^{} \,\,=\,\, U^\dagger_{\rm PMNS} U_{\nu N}^{} ~, \hspace{5ex}
\bar U_0^{} \,\,=\,\, U^\dagger_{\rm PMNS} U_0^{} ~, \hspace{5ex}
\bar U_\delta^{} \,\,=\,\, U^\dagger_{\rm PMNS} U_\delta^{} ~,
\end{eqnarray}
we have
\begin{eqnarray}
\bar U_{\nu N}^{} M_N^{} \bar U_{\nu N}^{\rm T} \,\,=\,\, - \hat m_\nu^{} ~.
\end{eqnarray}
Taking, for example,
\begin{eqnarray} \label{tUad}
\bar U_0^{} \,\,= \left( \begin{array}{ccc}
\bar a \,&\, \bar a \,&\, i\sqrt2\,\bar a \vspace{0.5ex} \\
\bar b & \bar b & i\sqrt2\,\bar b \vspace{0.5ex} \\ \bar c & \bar c & i\sqrt2\,\bar c
\end{array}\right) \!{\cal R} \,\,=\,\, U^\dagger_{\rm PMNS} U_0^a ~, \hspace{5ex}
\bar U_{\delta} \,\,=\,\, \left(\begin{array}{ccc}
\bar\delta_{11}^{} \,&\, \bar\delta_{12}^{} \,&\, \bar\delta_{13}^{} \vspace{0.5ex} \\
\bar\delta_{21}^{} & \bar\delta_{22}^{} & \bar\delta_{23}^{} \vspace{0.5ex} \\
\bar\delta_{31}^{} & \bar\delta_{32}^{} & \bar\delta_{33}^{} \end{array} \right) \!{\cal R} ~,
\end{eqnarray}
neglecting $\bar\delta^2_{ij}$ terms, and equating the elements of
\,$m\equiv\bar U_{\nu N}^{}M_N^{}\bar U_{\nu N}^{\rm T}$\,
to those of  \,$-\hat m_\nu^{}$,\, we then find
\begin{eqnarray}
m_{11}^{} &=& 2\bar a\left(\bar\delta_{11}^{}+\bar\delta_{12}^{}
+ i\sqrt2\,\bar\delta_{13}^{}\right) m_N^{} \,=\, -m_{\nu_1}^{} ~, \nonumber\\
m_{22}^{} &=& 2\bar b\left(\bar\delta_{21}^{}+\bar\delta_{22}^{}
+ i\sqrt2\,\bar\delta_{23}^{}\right) m_N^{} \,=\, -m_{\nu_2}~, \nonumber\\
m_{33}^{} &=& 2\bar c\left(\bar\delta_{31}^{}+\bar\delta_{32}^{}
+ i\sqrt2\,\bar\delta_{33}^{}\right) m_N^{} \,=\, -m_{\nu_3} ~, \nonumber\\
m_{12}^{} &=& m_{21}^{} \,=\,
\left[\bar b\left(\bar\delta_{11}^{}+\bar\delta_{12}^{}+i\sqrt2\,\bar\delta_{13}^{}\right)
+ \bar a\left(\bar\delta_{21}^{}+\bar\delta_{22}^{}
+ i\sqrt2\,\bar\delta_{23}^{}\right)\right]m_N^{}
\,=\, 0 ~, \nonumber \\
m_{13}^{} &=& m_{31}^{} \,=\,
\left[\bar c\left(\bar\delta_{11}^{}+\bar\delta_{12}^{}+i\sqrt2\,\bar\delta_{13}^{}\right)
+ \bar a\left(\bar\delta_{31}^{}+\bar\delta_{32}^{}
+ i\sqrt2\,\bar\delta_{33}^{}\right)\right]m_N^{}
\,=\, 0 ~, \nonumber \\
m_{23}^{} &=& m_{32}^{} \,=\,
\left[\bar c\left(\bar\delta_{21}^{}+\bar\delta_{22}^{}+i\sqrt2\,\bar\delta_{23}^{}\right)
+ \bar b\left(\bar\delta_{31}^{}+\bar\delta_{32}^{}
+ i\sqrt2\,\bar\delta_{33}^{}\right)\right]m_N^{}
\,=\, 0 ~.
\end{eqnarray}
Solving \,$m_{12}^{}=m_{23}^{}=0$\, first, we get
\begin{eqnarray}
m_{11}^{} &=& \frac{-2\bar a^2\, \bar\delta\, m_N^{}}{\bar b} ~, \hspace{5ex}
m_{22}^{} \,=\, +2\bar b\,\bar\delta\, m_N^{} ~, \hspace{5ex}
m_{33}^{} \,=\, \frac{-2\bar c^2\,\bar\delta\, m_N^{}}{\bar b} ~,
\nonumber \\ &&
m_{13}^{} \,=\, \frac{-2\bar a\,\bar c\,\bar\delta\, m_N^{}}{\bar b} ~, \hspace{5ex}
\bar\delta \,=\, \bar\delta_{21}^{}+\bar\delta_{22}^{}+i\sqrt2\,\bar\delta_{23}^{} ~.
\end{eqnarray}
It follows that, for $m_{13}^{}$ to vanish as well and nontrivial solutions be found, one of
the light-neutrino masses has to be zero.  Accordingly, we find two possible solutions:
\begin{eqnarray}
({\rm i}) \;\; \bar a &=& 0 , \hspace{5ex} \hat m_\nu^{} \,=\,
{\rm diag}\left(0,\,-1,\,\frac{\bar c^2}{\bar b^2}\right)2\bar b\,\bar\delta\,m_N^{} ~, \\
({\rm ii}) \;\; \bar c &=& 0, \hspace{5ex} \hat m_\nu^{} \,=\,
{\rm diag}\left(\frac{\bar a^2}{\bar b^2},\, -1,\, 0\right)2\bar b\,\bar\delta\,m_N^{} ~,
\end{eqnarray}
corresponding to the (i)~normal hierarchy $\bigl(m_{\nu_1}<m_{\nu_2}\ll m_{\nu_3}\bigr)$ and
(ii)~inverted hierarchy $\bigl(m_{\nu_3}\ll m_{\nu_1}<m_{\nu_2}\bigr)$ cases, respectively.
One can also obtain  \,$m_{\nu_2}=0$\,  solutions with  $\bar\delta$  replaced by
\,$\bar\delta_{11}+\bar\delta_{12}+i\sqrt2\,\bar\delta_{13}$,\, but
these solutions are phenomenologically unacceptable because
\,$\Delta m_{21}^2=m_{\nu_2}^2-m_{\nu_1}^2>0$\, according to solar neutrino
data~\cite{Schwetz:2008er},  implying that $m_{\nu_2}$ cannot be zero.

Before providing some numerical examples, we first note that
\,$m_{\nu_2}=-2\bar b\,\bar\delta\,m_N^{}$\, in both cases (i) and (ii).
This leads to
\begin{eqnarray}
{\rm(i)} \;\; \bar c^2 \,\,=\,\,
-\bar b^2\, \sqrt{\frac{\bigl|\Delta m_{31}^2\bigr|}{\Delta m_{21}^2}} ~, &\hspace{5ex}&
\bar\delta \,\,=\,\,
-50\; \frac{\sqrt{\Delta m_{21}^2}}{m_N^{}} \biggl(\frac{0.01}{\bar b}\biggr) ~,
\nonumber \\
{\rm(ii)} \;\; \bar a^2 \,\,=\,\, -\bar b^2\,
\sqrt{\frac{\bigl|\Delta m_{31}^2\bigr|}{\Delta m_{21}^2+\bigl|\Delta m_{31}^2\bigr|}} ~,
&\hspace{5ex}& \bar\delta \,\,=\,\,
-50\; \frac{\sqrt{\Delta m_{21}^2+\bigl|\Delta m_{31}^2\bigr|}^{\vphantom{\int}}}{m_N^{}}
\biggl(\frac{0.01}{\bar b}\biggr) ~.
\end{eqnarray}
Since our purpose is to illustrate that large elements in $U_{\nu N}$ are possible,
when necessary we take a relatively large number \,$\bar b=0.01$\, in numerical
calculations.
We also note that some of the $\bar\delta_{ij}$ do not play a role in the final determination
of the masses,  and so for the simplest solutions we can choose the nonzero $\bar\delta$'s
to be  \,$\bar\delta_{11}=-\bigl(\bar a/\bar b\bigr)\bar\delta_{21}$,\,
\,$\bar\delta_{31}=-\bigl(\bar c/\bar b\bigr)\bar\delta_{21}$,\, and
$\bar\delta_{21}$.

For demonstration, in the following we will present solutions using the central values of
\,$\Delta m_{21}^2 = \bigl(7.65^{+0.23}_{-0.20}\bigr)\times 10^{-5}$\,eV$^2$\, and
\,$\bigl|\Delta m_{31}^2\bigr| =\bigl(2.40^{+0.12}_{-0.11}\bigr)\times 10^{-3}$\,eV$^2$,\,
determined by a recent fit to global neutrino data~\cite{Schwetz:2008er}, and
$U_{\rm PMNS}$ in the tri-bimaximal form~\cite{Harrison:2002er}
\begin{eqnarray} \label{Utb}
U_{\rm tribi}^{} \,\,= \left( \begin{array}{ccc} \displaystyle
\mbox{$\frac{-2}{\sqrt6}$} & \mbox{$\frac{1}{\sqrt3}$} & 0 \vspace{0.5ex} \\ \displaystyle
\mbox{$\frac{1}{\sqrt6}$} & \mbox{$\frac{1}{\sqrt3}$} & \mbox{$\frac{1}{\sqrt2}$}
\vspace{0.5ex} \\ \displaystyle
\mbox{$\frac{1}{\sqrt6}$} \,&\, \mbox{$\frac{1}{\sqrt3}$} \,&\, \mbox{$\frac{-1}{\sqrt2}$}
\end{array} \right) .
\end{eqnarray}

Thus, using the relations in Eqs.~(\ref{tu}) and~(\ref{tUad}), we get
\begin{eqnarray} \label{i0}
{\rm(i)} \;\; a &=& 0.58\,\bar b ~, \hspace{5ex}
b \,=\, (0.58 + 1.7\, i)\bar b ~, \hspace{5ex}
c \,=\, (0.58 - 1.7\, i)\bar b ~, \nonumber \\
\delta_{11}^{} &=&
\frac{-2.5\,\rm GeV}{10^{12}\,\,\bar b\, m_N^{}} ~, \hspace{4ex}
\delta_{21}^{} \,=\,
\frac{(-2.5+7.3\,i)\,\rm GeV}{10^{12}\,\,\bar b\, m_N^{}} ~, \hspace{4ex}
\delta_{31}^{} \,=\,
\frac{(-2.5-7.3\,i)\,\rm GeV^{\vphantom{\sum}}}{10^{12}\,\,\bar b\, m_N^{}} ~, \hspace*{4ex}
\\ \nonumber \\ \label{ii0}
{\rm(ii)} \;\; a &=& (0.58 - 0.81\,i)\bar b ~, \hspace{5ex}
b \,=\, ( 0.58 + 0.41\,i)\bar b ~, \hspace{5ex}
c \,=\, ( 0.58 + 0.41\,i)\bar b ~, \nonumber \\ && \hspace*{-11ex}
\delta_{11}^{} \,=\,
\frac{(-1.4-2.0\,i)\,\rm GeV}{10^{11}\,\,\bar b\, m_N^{}} ~, \hspace{3ex}
\delta_{21}^{} \,=\,
\frac{(-1.4+1.0\,i)\,\rm GeV}{10^{11}\,\,\bar b\, m_N^{}} ~, \hspace{3ex}
\delta_{31}^{} \,=\,
\frac{(-1.4+1.0\,i)\,\rm GeV^{\vphantom{\sum}}}{10^{11}\,\,\bar b\,m_N^{}} ~, 
\end{eqnarray}
as possible solutions for  \,$U_{\nu N}^{}=U_0^{}+U_\delta^{}$,\, with \,$U_0^{}=U_0^a$\,
and the other $\delta$'s vanishing, corresponding to the (i)~normal hierarchy
$\bigl(m_{\nu_1}=0$,\, \,$m_{\nu_2}=0.00875$\,eV,\, \,$m_{\nu_3}=0.049$\,eV$\bigr)$ and
(ii)~inverted hierarchy
$\bigl(m_{\nu_1}=0.049$\,eV,\, \,$m_{\nu_2}=0.0498$\,eV,\, \,$m_{\nu_3}=0\bigr)$ cases,
respectively.  These examples show indeed that large mixing of light and heavy neutrinos can
be found and at the same time small neutrino masses are maintained.

In Appendix~\ref{upmns} we also provide, using the empirical $U_{\rm PMNS}$, the counterparts
of the numbers in Eqs.~(\ref{i0}) and~(\ref{ii0}), showing that solutions can also be found in
the more general case. As expected, the two sets of results are numerically similar to each other.

\subsection{Solutions with all light-neutrino masses being nonzero}

In the previous examples, one of the light neutrinos is massless.
We find that if one allows another correction matrix, $U_{\alpha\beta\gamma}$, whose elements
are of order $\bigl[(a,b,c)\delta_{ij}\bigr]{}^{\!1/2}$, one can obtain solutions for both
the normal and inverted hierarchies, with all the three light-neutrino masses being nonzero.

For instance, we consider
\begin{eqnarray} \label{tUdd}
\bar U_0^{} \,\,= \left( \begin{array}{lll} \vspace{0.5ex}
0 & \bar a & i\bar a \\ 0 & \bar b & i\bar b \vspace{0.5ex} \\ 0 & \bar c & i\bar c
\end{array}\right) \!{\cal R} \,\,=\,\, U^\dagger_{\rm PMNS} U_0^d ~, \hspace{5ex}
\bar U_{\alpha\beta\gamma} \,\,= \left(\begin{array}{lll} \bar\alpha & 0 & 0 \vspace{0.5ex} \\
\bar\beta & 0 & 0 \vspace{0.5ex} \\ \bar\gamma & 0 & 0 \end{array}\right) \!{\cal R}
=\,\, U^\dagger_{\rm PMNS} U_{\alpha\beta\gamma}^{} ~.
\end{eqnarray}
with  \,$\bar b=0$\,  and  $\bar U_\delta$ as in Eq.~(\ref{tUad}).
Thus in this case  \,$U_{\nu N}^{}=U_0^d+U_{\alpha\beta\gamma}^{}+U_\delta^{}$.\,
Since $\bar\alpha$, $\bar\beta$ and $\bar\gamma$ are of order
$\bigl[(\bar a,\bar c)\bar\delta_{ij}\bigr]{}^{\!1/2}$, one should keep $\bar\alpha^2$,
$\bar\beta^2$, and $\bar\gamma^2$ terms in the calculation, neglecting
$\bar\delta_{ij}\bar\delta_{kl}$ and $(\bar\alpha,\bar\beta,\bar\gamma)\bar\delta_{kl}$ terms.
Upon equating  \,$m=\bar U_{\nu N}^{}M_N^{}\bar U_{\nu N}^{\rm T}$\,
to \,$-\hat m_\nu^{}$\, as before,  we arrive at
\begin{eqnarray}
m_{11}^{} &=&
\left[2\bar a\,\Bigl(\bar\delta_{12}^{}+i\bar\delta_{13}^{}\Bigr)+\bar\alpha^2\right]m_N^{}
\,=\, -m_{\nu_1} ~,  \nonumber \\
m_{22}^{} &=& \bar\beta^2\, m_N^{} \,=\, -m_{\nu_2} ~, \nonumber \\
m_{33}^{} &=&
\left[2\bar c\,\Bigl(\bar\delta_{32}^{}+i\bar\delta_{33}^{}\Bigr)+\bar\gamma^2\right]m_N^{}
\,=\, -m_{\nu_3} ~, \nonumber\\
m_{12}^{} &=& m_{21}^{} \,=\,
\left[\bar a\,\Bigl(\bar\delta_{22}^{}+i\bar\delta_{23}^{}\Bigr)
+ \bar\alpha\,\bar\beta\right]m_N^{} \,=\, 0 ~, \nonumber \\
m_{13}^{} &=& m_{31}^{} \,=\,
\left[\bar a\,\Bigl(\bar\delta_{32}^{}+i\bar\delta_{33}^{}\Bigr)
+ \bar c\left(\bar\delta_{12}^{}+i\bar\delta_{13}^{}\right)
+ \bar\alpha\,\bar\gamma \right] m_N^{} \,=\, 0 ~, \nonumber \\
m_{23}^{} &=& m_{32}^{} \,=\,
\left[\bar c\,\Bigl(\bar\delta_{22}^{}+i\bar\delta_{23}^{}\Bigr)
+ \bar\beta\,\bar\gamma \right]m_N^{} \,=\, 0 ~,
\end{eqnarray}
leading to
\begin{eqnarray}
\hat m_\nu^{} \,\,=\,\, {\rm diag}\left( \frac{\bar a^2\,\bar\gamma^2}{\bar c^2}
+ \frac{2\bar a^2}{\bar c}\bigl(\bar\delta_{32}^{}+i\bar\delta_{33}^{}\bigr),\, -\bar\beta^2,\,
-\bar\gamma^2-2\bar c\left(\bar\delta_{32}^{}+i\bar\delta_{33}^{}\right) \right) m_N^{} ~,
\end{eqnarray}
\begin{eqnarray}
\bar\alpha \,\,=\,\, \frac{\bar a\,\bar\gamma}{\bar c} ~, \hspace{5ex}
\bar\delta_{12}^{} \,\,=\,\, \frac{-\bar a\,\bar\gamma^2
- \bar a\,\bar c\left(\bar\delta_{32}^{}+i\bar\delta_{33}^{}\right)
- i\bar c^2\,\bar\delta_{13}^{}}{\bar c^2} ~, \hspace{5ex}
\bar\delta_{22}^{} \,\,=\,\, -i\bar\delta_{23}^{} - \frac{\bar\beta\,\bar\gamma}{\bar c} ~.
\end{eqnarray}
The resulting solutions involve simple expressions:
\begin{eqnarray}
&& \bar a^2 \,\,=\,\, -\bar c^2\, \sqrt{\frac{m^2_{\nu_2}-\Delta m_{21}^2}
{m^2_{\nu_2}-\Delta m_{21}^2+\Delta m_{31}^2}} ~, \hspace{5ex}
\bar\beta^2 \,\,=\,\, -{m_{\nu_2}\over m_N^{}} ~, \nonumber \\
\bar\gamma^2 &\,=\,& \frac{-1}{m_N^{}}\sqrt{m^2_{\nu_2}-\Delta m_{21}^2+\Delta m_{31}^2} ~,
\hspace{5ex} \bar\delta_{12}^{} \,=\, -\frac{\bar a\,\bar\gamma^2}{\bar c^2} ~, \hspace{5ex}
\bar\delta_{22}^{} \,\,=\,\, -\frac{\bar\beta\,\bar\gamma}{\bar c} ~,
\end{eqnarray}
with the other $\bar\delta_{ij}$ having been set to zero.

For numerical illustrations, we again adopt  \,$U_{\rm PMNS}=U_{\rm tribi}$\,  and
the central values of neutrino data quoted above and choose \,$m_{\nu_2}=0.1$\,eV.\,
Employing the relations in Eqs.~(\ref{tu}) and~(\ref{tUdd}) between the barred and unbarred
quantities,  for  \,$U_{\nu N}^{}=U_0^{}+U_{\alpha\beta\gamma}^{}+U_\delta^{}$,\,  with
\,$U_0^{}=U_0^d$,\,  we obtain as possible solutions
\begin{eqnarray} \label{i0'}
{\rm(i)} \;\; a &=& -0.82\,\bar a ~, \hspace{5ex}
b \,=\, (0.41 + 0.75\,i)\bar a ~, \hspace{5ex}
c \,=\, (0.41 - 0.75\,i)\bar a ~, \nonumber \\ \vphantom{\sum_|^|}
\alpha  &=&  \frac{ 8.1-5.8\,i}{10^6\,\sqrt{m_N^{}/\rm GeV}} ~, \hspace{4ex}
\beta  \,=\, \frac{-4.1- 13\,i}{10^6\,\sqrt{m_N^{}/\rm GeV}} ~, \hspace{4ex}
\gamma \,=\, \frac{-4.1+1.7\,i}{10^6\,\sqrt{m_N^{}/\rm GeV}} ~, \\
\delta_{12}^{} &=&   \frac{( 8.1-5.8\,i)\,\rm GeV}{10^{11}\,\bar a\,m_N^{}} ~, \hspace{4ex}
\delta_{22}^{} \,=\, \frac{(-4.1-5.8\,i)\,\rm GeV}{10^{11}\,\bar a\,m_N^{}} ~, \hspace{4ex}
\delta_{32}^{} \,=\, \frac{(-4.1-5.8\,i)\,\rm GeV}{10^{11}\,\bar a\,m_N^{}} \nonumber
\end{eqnarray}
in the normal-hierarchy case
$\bigl(m_{\nu_1}=0.0996$\,eV,\, \,$m_{\nu_2}=0.1$\,eV,\, \,$m_{\nu_3}=0.111$\,eV$\bigr)$ and
\begin{eqnarray} \label{ii0'}
{\rm(ii)} \;\; a &=& -0.82\,\bar a ~, \hspace{5ex}
b \,=\, ( 0.41+0.66\,i)\bar a ~, \hspace{5ex}
c \,=\, ( 0.41-0.66\,i)\bar a ~, \nonumber \\ \vphantom{\sum_|^|}
\alpha  &=&  \frac{ 8.1-5.8\,i}{10^6\,\sqrt{m_N^{}/\rm GeV}} ~, \hspace{4ex}
\beta  \,=\, \frac{-4.1- 12\,i}{10^6\,\sqrt{m_N^{}/\rm GeV}} ~, \hspace{4ex}
\gamma \,=\, \frac{-4.1+0.8\,i}{10^6\,\sqrt{m_N^{}/\rm GeV}} ~, \\
\delta_{12}^{} &=&   \frac{( 8.1-5.8\,i)\,\rm GeV}{10^{11}\,\bar a\,m_N^{}} ~, \hspace{4ex}
\delta_{22}^{} \,=\, \frac{(-4.1-5.8\,i)\,\rm GeV}{10^{11}\,\bar a\,m_N^{}} ~, \hspace{4ex}
\delta_{32}^{} \,=\, \frac{(-4.1-5.8\,i)\,\rm GeV}{10^{11}\,\bar a\,m_N^{}} \nonumber
\end{eqnarray}
in the inverted-hierarchy case
$\bigl(m_{\nu_1}=0.0996$\,eV,\, \,$m_{\nu_2}=0.1$\,eV,\, \,$m_{\nu_3}=0.0867$\,eV$\bigr)$,
with the other $\delta$'s vanishing.

We have again collected in Appendix~\ref{upmns} the corresponding numbers obtained using
the empirical $U_{\rm PMNS}$, demonstrating that solutions can also be found in the general case.
The resulting numbers in Eqs.~(\ref{i'}) and~(\ref{ii'}) are as expected similar to those
in Eqs.~(\ref{i0'}) and~(\ref{ii0'}).

\subsection{Some implications for probing type-I seesaw at the LHC}

We have seen above that the elements of $U_{\nu N}$ can be large and simultaneously
satisfy the constraints from the tiny neutrino masses.
There are two other classes of processes which also provide constraints on the $U_{\nu N}$
elements.  The first involves neutral currents conserving lepton flavor and can be used to
test deviations from the SM predictions for electroweak precision observables~\cite{ewpd}.
They have been measured mainly at LEP~\cite{pdg}, and for type-I seesaw the bounds extracted
from the data are  \,{\footnotesize$\sum$}$_i|(U_{\nu N})_{1i}|^2\le 0.0030$,\,
\,{\footnotesize$\sum$}$_i|(U_{\nu N})_{2i}|^2\le 0.0032$,\, and
\,{\footnotesize$\sum$}$_i|(U_{\nu N})_{3i}|^2\le 0.0062$\,~\cite{delAguila:2008cj,ewpd}.
The second class of processes consists of FCNC transitions in the charged-lepton sector.
Although in type-I seesaw there are no FCNC processes involving ordinary charged leptons at
tree level, loop-induced ones can occur, such as the radiative decays  \,$\mu\to e\gamma$,\,
\,$\tau\to e\gamma$,\, and \,$\tau\to\mu\gamma$.\,
For type-I seesaw, the constraints determined from the measurements of these FCNC transitions are
\,$\big|${\footnotesize$\sum$}$_i^{}(U_{\nu N})_{1i}^{}(U_{\nu N})^*_{2i}\big|\le 0.0001$,\,
\,$\big|${\footnotesize$\sum$}$_i^{}(U_{\nu N})_{1i}^{}(U_{\nu N})^*_{3i}\big|\le 0.01$,\, and
\,$\big|${\footnotesize$\sum$}$_i^{}(U_{\nu N})_{2i}^{}(U_{\nu N})^*_{3i}\big|\le
0.01$\,~\cite{ewpd,typeI_fcnc},
obviously the first one being very restrictive.
In view of these bounds, we find that for the solutions given in Eqs.~(\ref{tUad}) and
(\ref{tUdd}) the elements of $U_{\nu N}$ can be as large as~0.01.
However, there are other types of solutions which we obtain in the next section for
type-III seesaw that also work for type-I seesaw and can better evade these constraints.
Choosing, for example, $U_0$ of the form $U_0^e$ given in Eq.~(\ref{u0e}), we can easily
see that all the constraints above are satisfied by the nonzero elements of $U_{\nu N}$
having size up to~0.04.

We can then first take \,$\bar b\sim0.01$\, in the cases with one of the light neutrinos being
massless or  \,$\bar a\sim0.01$\, in the cases with all of the light-neutrino masses being nonzero.
With this choice, the elements of $U_{\nu N}$ in the examples treated above are at most of
order~0.01.  We can now consider how such numbers translate into the production of $N$ at the LHC.
Specifically, we concentrate on the production channel \,$q\bar q'\to W^*\to l N$,\,
as it involves a light charged lepton $l$, which makes the signal more detectable.

We have explored the cross section $\sigma$ for \,$p p\to l N X$\, arising from
\,$q\bar q'\to W^*\to l N$\,  at $pp$ center-of-mass energy of  \,$\sqrt s=14$\,TeV.\,
For the parton distribution functions, we employ those provided by Ref.~\cite{mstw}.
The resulting plot as a function of the $N$ mass is already shown in Fig.~\ref{csplot}.
It indicates that, with $U_{\nu N}$ elements of order~0.01, heavy neutrinos having
masses as large as \,$m_N^{}=115$\,GeV\, can be produced at a cross section of at least 1\,fb.
With the $U_{\nu N}$ elements allowed to have the larger size of~0.04 instead, still consistent
with the experimental bounds, cross-section values higher than~1\,fb can be reached for
$m_N^{}$ up to 250\,GeV.  Moreover, with 100\,fb$^{-1}$ of integrated luminosity,
the production of over 3000 heavy neutrinos having a 100-GeV mass is possible,
the number of events dropping to a few for \,$m_N^{}=600$\,GeV.\,
A recent analysis including background estimates suggests that neutrinos of masses up
to 150\,GeV can be observed at the LHC with 30\,fb$^{-1}$ luminosity~\cite{del Aguila:2007em}.

\begin{figure}[!t]
\includegraphics[width=171mm]{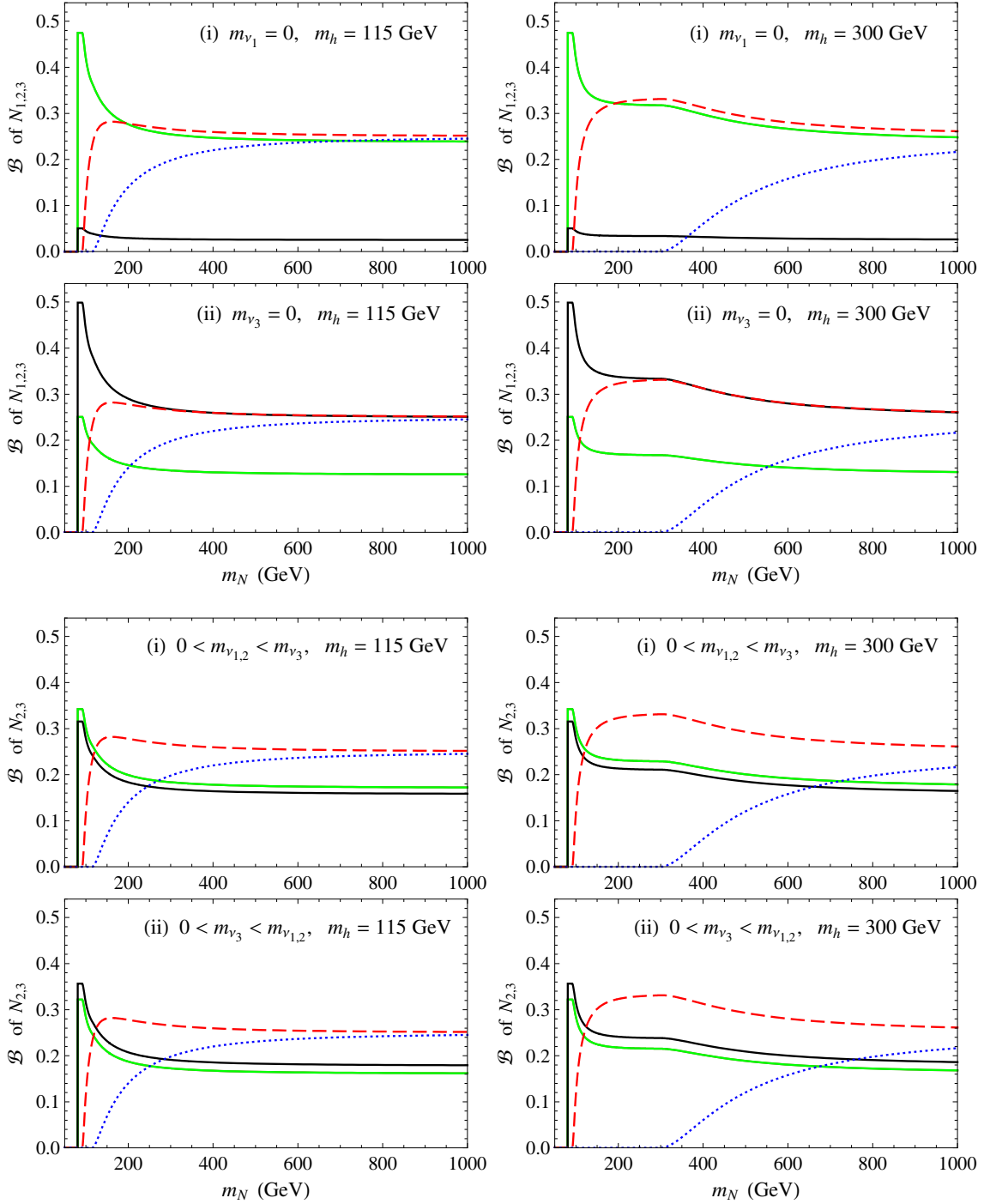}
\caption{Branching ratios $\cal B$ of \,$N\to l W$ (solid curves), \,$N\to\nu Z$ (dashed red curves),
and \,$N\to\nu h$ (dotted blue curves)\, as functions of the $N$ mass for the sample solutions of
$U_{\nu N}$ found in type-I seesaw and Higgs masses \,$m_h^{}=115$ and 300\,GeV.\,
The top (bottom) four plots correspond to the examples with one (none) of the light neutrinos
being massless, in either the normal~(i) or inverted~(ii) hierarchy case.
Each black solid curve belongs to \,$N\to e W$,\, whereas a green (gray) solid curve represents
both \,$N\to\mu W$ and $N\to\tau W$,\, which are equal in branching ratio.\label{brplot1}}
\end{figure}

If the LHC does observe the heavy neutrinos, it will be interesting to study their decay rates
and branching ratios to gain more information about the light-heavy and/or light-neutrino
mixing, as well as the light-neutrino masses.
As the expressions collected in Appendix~\ref{widths} indicate, the rates can reveal some
information about the size of the light-heavy mixing parameterized by $U_{\nu N}$.
If this mixing is small, $U_{\nu N}$ will be related to $\hat m_\nu^{}$ and $U_{\rm PMNS}$
by Eq.~(\ref{UmU}), and so the information on $U_{\nu N}$ may in turn reveal something
about the light-neutrino masses and mixing.
In the case of large light-heavy mixing treated in this paper, the dominant part of $U_{\nu N}$
is decoupled from the light-neutrino masses.
In that case, one can nevertheless still learn something about the neutrino masses, besides
the light-heavy mixing, by examining the branching ratios.
Their expressions for the dominant decays can be derived from Appendix~\ref{widths}.
We have plotted them in Fig.~\ref{brplot1} for the examples given in this section.
We remark that the curves belonging to the $\nu Z$ and $\nu h$ modes have been obtained
after summing over contributions with $\nu_{1,2,3}^{}$ in the final states, and that
in the top four plots the different heavy neutrinos $N_{1,2,3}$ have the same branching
ratios, whereas in the bottom four plots only $N_{2,3}$ have the same branching ratios, with
the $N_1$ branching ratios not shown due to its decay widths being negligible.
The graphs in Fig.~\ref{brplot1} illustrate that different patterns of the values of
the $U_{\nu N}$ elements generally translate into different patterns of the branching ratios.
Moreover, studying the branching ratios could also uncover what type of mass hierarchy
the light neutrinos might have.

\section{Type-III seesaw and FCNC involving light charged leptons}

Most of the analysis of the preceding section can be applied to the seesaw scenario of type~III.
However, in the type~III there are also new heavy charged-leptons associated with
the heavy neutrinos~\cite{Foot:1988aq}.
This new feature leads to some additional interesting implications concerning the possibility
of large mixing between the light and heavy neutrinos, which we deal with here.

In type-III seesaw the SM-singlet neutrinos in type-I seesaw are replaced by weak-SU$(2)_{\rm L}$
triplets of right-handed heavy leptons having zero hypercharge~\cite{Foot:1988aq}.
The component fields of each triplet $\Sigma$ and its charge conjugate
\,$\Sigma^c=C\bar\Sigma^{\rm T}$\, are
\begin{eqnarray}
\Sigma \,\,= \left(\begin{array}{cc} N^0/\sqrt2 & E^+ \vspace{0.5ex} \\
E^- & -N^0/\sqrt2 \end{array} \right), \hspace{5ex}
\Sigma^c \,\,= \left( \begin{array}{cc} N^{0c}/\sqrt2 & E^{-c} \vspace{0.5ex} \\
E^{+c} & -N^{0c}/\sqrt2 \end{array} \right) ,
\end{eqnarray}
and the renormalizable Lagrangian for each $\Sigma$ is given by
\begin{eqnarray} \label{Liii0}
{\cal L}_{\rm III}^{} \,\,=\,\,
{\rm Tr}\bigl(\bar\Sigma i\!\!\not{\!\!D}\Sigma\bigr)
- \mbox{$\frac{1}{2}$}\, {\rm Tr}\Bigl(\bar\Sigma M_\Sigma^{}\Sigma^c
+ \overline{\Sigma^{{}^{\scriptstyle c}}} M_\Sigma^* \Sigma\Bigr)
- \sqrt2\, \tilde H^\dagger\bar\Sigma Y_\Sigma^{} L_L^{}
- \sqrt2\, \bar l_L^{} Y_\Sigma^\dagger \Sigma \tilde H ~,
\end{eqnarray}
where  $D_\mu$ is a covariant derivative involving the weak gauge bosons, $M_\Sigma^{}$
the mass matrix of the triplet, and $Y_\Sigma$ the Yukawa-coupling matrix.
Defining  \,$E=\bigl(E_R^+\bigr){}^c+E_R^-$\, and removing the would-be Goldstone bosons $\eta$
and $\phi^\pm$, one can rewrite ${\cal L}_{\rm III}$ as
\begin{eqnarray} \label{Liii}
{\cal L}_{\rm III}^{} &=& \bar E i\!\!\not{\!\partial} E
+ \bar N^0_R  i\!\!\not{\!\partial} N^0_R - \bar E M_\Sigma^{} E
- \mbox{$\frac{1}{2}$}\Bigl[\bar N^0_R M_\Sigma^{}\bigl(N_R^0\bigr)^c\,+\,{\rm H.c.}\Bigr]
\nonumber \\ && \!\! +\,\,
g \left[\bar N^0_R  \not{\!\!W}^+ E_R^{}
+ \overline{\bigl(N_R^0\bigr)^{\!c}} \not{\!\!W}^+ E_L^{} \,+\, {\rm H.c.} \right]
- g\, \bar E \not{\!\!W}_{\!\!3}^{} E
\nonumber \\ && \!\! -\,\,
\Bigl[ \mbox{$\frac{1}{\sqrt2}$}(v+h)\bar N^0_R Y_\Sigma^{}\nu_L^{}
+ (v+h)\bar E Y_\Sigma^{} l_L^{} \,+\, {\rm H.c.} \Bigr] ~,
\end{eqnarray}
where  \,$W_3^\mu=-s_{\rm w}^{}A^\mu+c_{\rm w}^{}Z^\mu$,\,
with \,$s_{\rm w}^{}=\sin\theta_{\rm W}^{}$,\,  is the usual linear combination of
the photon and $Z$-boson fields,  \,$N_R=N$,\, and  \,$E_{L,R}=P_{L,R}E$,\, with
\,$P_{L,R}=\frac{1}{2}(1\mp\gamma_5^{})$.\,

From this Lagrangian, one can easily identify the terms for the lepton masses.
For three generations, the neutrino-mass matrix has the seesaw form given by
\begin{eqnarray} \label{Lm'}
{\cal L}_{\rm mass}' \,\,=\,\,
-\mbox{$\frac{1}{2}$} \Bigl(\overline{\bigl(\nu_L^{}\bigr)^c} \hspace{3ex} \bar N^0 \Bigr)
\left(\begin{array}{cc} 0 & Y_\Sigma^{\rm T} v/\sqrt2 \vspace{0.5ex} \\
Y_\Sigma^{} v/\sqrt2 & M_\Sigma^{} \end{array} \right)
\left(\begin{array}{c} \nu_L^{} \vspace{0.5ex} \\ N^{0c} \end{array} \right)
\,+\,\, {\rm H.c.} ~,
\end{eqnarray}
and the charged associates of the heavy neutrinos mix with the SM charged leptons resulting in
the mass-matrix term
\begin{eqnarray}
{\cal L}_{\rm mass}'' \,\,=\,\, -\Bigl(\bar l_R^{} \hspace{3ex} \bar E_R^{}\Bigr)
\left(\begin{array}{cc} m_l^{} & 0 \vspace{0.5ex} \\
Y_\Sigma^{}v & M_\Sigma^{} \end{array}\right)
\left( \begin{array}{c} l_L^{} \vspace{0.5ex} \\ E_L^{} \end{array} \right)
\,+\,\, {\rm H.c.} ~,
\end{eqnarray}
where $Y_\Sigma$, $M_\Sigma$, and $m_l^{}$ are now 3$\times$3 matrices and
$\nu$, $N$, $l$, and $E$ are 3$\times$1 (column) matrices.
Without loss of generality, from this point on, we work in the basis in which  $M_\Sigma^{}$
and $m_l^{}$ are already real and diagonalized:
\,$M_{\Sigma}^{}={\rm diag}\bigl(1/r_1^{},1/r_2^{},1/r_3^{}\bigr)m_N^{}$,
with \,$r_i^{}=m_N^{}/M_i^{}$.
One can diagonalize the (6$\times$6) mass matrices by transforming from the weak
eigenstates to mass eigenstates using the relations
\begin{eqnarray}
\left(\begin{array}{c} \nu_L^{} \vspace{0.5ex} \\ N^{0c} \end{array}\right) \,\,=\,\,
U \left(\begin{array}{c} \nu_{mL}^{} \vspace{0.5ex}\\ N_{mL}^{} \end{array}\right), \hspace{5ex}
\left(\begin{array}{c} l_{L,R}^{} \vspace{0.5ex} \\ E_{L,R}^{} \end{array}\right) \,\,=\,\,
U_{L,R}^{} \left(\begin{array}{c} l_{mL,mR}^{} \\ E_{mL,mR}^{} \end{array}\right) ,
\end{eqnarray}
where $U_{L,R}$ and $U$ are (3+3)-by-(3+3) unitary matrices if three triplets are present
and can be expressed as
\begin{eqnarray}
U \,\,=\,\, \left(\begin{array}{cc} U_{\nu\nu}^{} & U_{\nu N}^{} \vspace{0.5ex} \\
U_{N\nu}^{} & U_{NN}^{} \end{array}\right) , \hspace{5ex}
U_L^{} \,\,=\,\, \left(\begin{array}{cc} U_{Lll}^{} & U_{LlE}^{} \vspace{0.5ex} \\
U_{LE l}^{} & U_{L EE} \end{array}\right) , \hspace{5ex}
U_{R} \,\,=\,\, \left(\begin{array}{cc} U_{Rll}^{} & U_{R lE}^{} \vspace{0.5ex} \\
U_{RE l}^{} & U_{R EE}^{} \end{array}\right) .
\end{eqnarray}
From the diagonalization calculation, we derive to second order in
\,$Y_\Sigma^{}v M^{-1}_\Sigma$\, and/or \,$m_l^{}M^{-1}_\Sigma$\,
\begin{eqnarray} \label{u}
U_{\nu\nu}^{} \,\,=\,\, \Bigl(1-\mbox{$\frac{1}{2}$}\,\epsilon\Bigr)U_{\rm PMNS}^{} ~, &&
U_{\nu N}^{} \,\,=\,\, \mbox{$\frac{1}{\sqrt2}$}\,Y^\dagger_\Sigma M^{-1}_\Sigma v ~,
\nonumber \\
U_{N \nu}^{} \,\,=\,\,
\mbox{$\frac{-1}{\sqrt2}$}\,M^{-1}_\Sigma Y_\Sigma^{} U_{\nu\nu}^{}\,v ~, &&
U_{NN}^{} \,\,=\,\, 1-\mbox{$\frac{1}{2}$}\,\epsilon' ~, \\
U_{Lll}^{} \,\,=\,\, 1- \epsilon ~, \hspace{5ex}
U_{LlE}^{} \,\,=\,\, Y^\dagger_\Sigma M^{-1}_\Sigma v ~, & \phantom{\int_\sum^{\sum^|}} \;\; &
U_{LEl}^{} \,\,=\,\, - M^{-1}_\Sigma Y_\Sigma^{} v ~, \hspace{5ex}
U_{LEE}^{} \,\,=\,\, 1-\epsilon' ~, \;\;\; \\
U_{Rll}^{} \,\,=\,\, 1 ~, \hspace{5ex}
U_{RlE}^{} \,\,=\,\, m_l^{} Y^\dagger_\Sigma M^{-2}_\Sigma v ~, & &
U_{REl}^{} \,\,=\,\, - M^{-2}_\Sigma Y_\Sigma^{} m_l^{} v ~, \hspace{5ex}
U_{REE}^{} \,\,=\,\, 1 ~,
\end{eqnarray}
where
\begin{eqnarray}
\epsilon \,\,\equiv\,\, \mbox{$\frac{1}{2}$}\,Y^\dagger_\Sigma M^{-2}_\Sigma Y_\Sigma^{} v^2
\,\,=\,\, U_{\nu N}^{} U_{\nu N}^{\dagger} ~, \hspace{5ex}
\epsilon' \,\,\equiv\,\,
\mbox{$\frac{1}{2}$}\,M^{-1}_\Sigma Y_\Sigma^{} Y^\dagger_\Sigma M^{-1}_\Sigma v^2 \,\,=\,\,
U_{\nu N}^{\dagger} U_{\nu N}^{} ~.
\end{eqnarray}

From ${\cal L}_{\rm III}$ in the mass-eigenstate basis, one can then write down the relevant
terms describing the interactions of the heavy leptons $N$ and $E$ with the SM gauge and Higgs
bosons.  The terms for the interactions of $N$ are the same as those in the type-I seesaw
covered earlier.  The corresponding interactions of $E$, at leading order, are described by
\begin{eqnarray} \label{LE}
{\cal L}_E^{} &=&
- g\,\overline{\bigl(\nu_{mL}^{}\bigr)^{\!c}}\not{\!\!W}^+U_{\rm PMNS}^{\rm T}U_{\nu N}^*E_{mR}^{}
\,+\, \frac{g}{\sqrt2\,c_{\rm w}}\, \bar l_{mL}^{}\not{\!\!Z} U_{\nu N}^{} E_{mL}^{}
\nonumber \\ && \!\! -\,\,
\frac{g}{\sqrt2\,m_W^{}}\, \bar l_{mL}^{}\, U_{\nu N}^{} M_{\Sigma}^{} E_{mR}^{} h
~+~ {\rm H.c.}~.
\end{eqnarray}
Using these couplings, one can derive the expressions for the rates of the dominant decay modes
of $N$ and $E$, which we have collected in Appendix~\ref{widths}.
In terms of the mass eigenstates, ${\cal L}_{\rm III}$ also contains the interactions of
the light charged leptons with the $Z$ and Higgs bosons.  To first order in~$\epsilon$, one has
\begin{eqnarray} \label{Ll}
{\cal L}_l^{} &=& \frac{g}{c_{\rm w}^{}}\,\bar l_m^{}\gamma^\mu
\left[ \Bigl( -\mbox{$\frac{1}{2}$}+s_{\rm w}^2-\epsilon\Bigr)P_L^{}
+ s_{\rm w}^2\,P_R^{}\right]l_m^{}\,Z_\mu^{}
\nonumber \\ && \!\! +\,\,
\frac{g}{2m_W^{}}\,\bar l_m^{} \left[ m_l^{}(3\epsilon-1)P_L^{}
+ (3\epsilon-1)m_l^{}\,P_R^{} \right] l_m^{}\, h ~.
\end{eqnarray}
It is evident from this Lagrangian that the off-diagonal elements of~$\epsilon$ are new sources
of tree-level FCNC's in the charged-lepton sector.
There are also tree-level FCNC's involving the light neutrinos arising from their couplings to
$Z$ and $h$ in Eq.~(\ref{L'}) in type-I seesaw and the corresponding ones in type-III seesaw,
but these interactions are difficult to observe and thus will not be discussed further.
We note that the results in Eqs.~(\ref{u})-(\ref{Ll}) agree with those given in
Ref.~\cite{Abada:2008ea}.

The effects of the off-diagonal elements of~$\epsilon$ on a~variety of processes involving charged
leptons have been evaluated in Refs.~\cite{Abada:2007ux,Abada:2008ea,He:2009tf,Arhrib:2009xf}.
The FCNC processes studied include  \,$l_i\to l_j\bar l_k l_l$,\, \,$l_i\to l_j\gamma$,\,
\,$Z\to l_i\bar l_j$,\, \,$\mu$-$e$ conversion in atomic nuclei,  \,$\tau\to M l$,\,
\,$M\to l\bar l'$,\, \,$M\to M'l\bar l'$,\, and muonium-antimuonium oscillation,
with $M$ denoting a meson.
Some of the existing experimental data on these transitions yield strict bounds on
the off-diagonal elements of~$\epsilon$.  The strongest constraint on $\epsilon_{12}^{}$ was
found to be \,$|\epsilon_{12}|<1.7\times 10^{-7}$\,  from $\mu$-$e$ conversion in atomic
nuclei~\cite{Abada:2008ea}.  Lepton-flavor violating processes involving the $\tau$ provide
very stringent bounds on~$\epsilon_{i3}^{}$.
Specifically, \,$\tau\to\pi^0 e$\, and  \,$\tau\to\mu\bar\mu\mu$\, yield
\,$|\epsilon_{13}^{}|<4.2\times 10^{-4}$\,  and  \,$|\epsilon_{23}^{}|<4.9\times 10^{-4}$,
respectively~\cite{He:2009tf}.

With these constraints from FCNC transitions, plus constraints from the tiny neutrino
masses, it is of interest to explore if the type-III seesaw scenario can still generate
the mixing of light and heavy neutrinos, as parameterized by $U_{\nu N}$, that is
large enough to be measurable at the LHC, as in the type-I case.
It is also of interest to find solutions for $U_{\nu N}$ that do not give rise to certain
$\epsilon_{ij}^{}$ that are too suppressed so as to make the corresponding FCNC processes
too small to probe.

For our first trial solutions, we use the results from the previous section.  Upon comparing
the heavy-neutrino sector in type-III seesaw above with that in type-I seesaw in the preceding
section, one can see that they are very similar.  Consequently, we can directly import here for
type-III seesaw the numerical solutions given in Eqs.~(\ref{i0}), (\ref{ii0}), (\ref{i0'}),
and~(\ref{ii0'}), which we subsequently use to obtain the matrix
\,$\epsilon=U_{\nu N}^{}U_{\nu N}^\dagger$.\,
Thus, neglecting the small corrections $U_\delta$ and $U_{\alpha\beta\gamma}$,  we have
for the cases with one of the light-neutrino masses vanishing
\begin{eqnarray} \label{eps1}
{\rm(i)} \;\; \epsilon \,\,=\, \left(\begin{array}{ccc}
 0.33 &\, 0.33 - 0.97\,i \,&\, 0.33 + 0.97\,i \vspace{0.5ex} \\
 0.33 + 0.97\,i \,&  3.1   &   -2.5 +1.9\,i \vspace{0.5ex} \\
 0.33 - 0.97\,i   & -2.5-1.9 i  & 3.1
\end{array}\right) \bigl|\bar b\bigr|^2\, \bigl(r_1^{} + r_2^{} + 2 r_3^{}\bigr) ~,
\end{eqnarray}
\begin{eqnarray} \label{eps2}
{\rm(ii)} \;\; \epsilon \,\,=\, \left(\begin{array}{ccc}
 0.99 &\, 0.01-0.70\,i \,&\, 0.01-0.70\,i \vspace{0.5ex} \\
 0.01+0.70\,i \,& 0.50  & 0.50 \vspace{0.5ex} \\
 0.01+0.70\,i   & 0.50  & 0.50
\end{array}\right) \bigl|\bar b\bigr|^2\, \bigl(r_1^{} + r_2^{} + 2 r_3^{}\bigr) ~,
\end{eqnarray}
corresponding to Eqs.~(\ref{i0}) and (\ref{ii0}), respectively, and for the cases
with all the light-neutrino masses being nonzero
\begin{eqnarray} \label{eps3}
{\rm(i)} \;\; \epsilon \,\,=\, \left(\begin{array}{ccc}
 0.67 &\, -0.33+0.61\,i \,&\, -0.33-0.61\,i \vspace{0.5ex} \\
-0.33-0.61\,i \,&  0.72   &   -0.39+0.61\,i \vspace{0.5ex} \\
-0.33+0.61\,i   & -0.39-0.61\,i  & 0.72
\end{array}\right) |\bar a|^2\, \bigl(r_2^{} + r_3^{}\bigr) ~,
\end{eqnarray}
\begin{eqnarray} \label{eps4}
{\rm(ii)} \;\; \epsilon \,\,=\, \left(\begin{array}{ccc}
 0.67 &\, -0.33+0.54\,i \,&\, -0.33-0.54\,i \\
-0.33-0.54\,i \,& 0.60    &   -0.27+0.54\,i \\
-0.33+0.54\,i   & -0.27-0.54\,i  & 0.60
\end{array}\right) |\bar a|^2\, \bigl(r_2^{} + r_3^{}\bigr) ~,
\end{eqnarray}
corresponding to Eqs.~(\ref{i0'}) and~(\ref{ii0'}), respectively.

In all these four cases, we notice that $\epsilon_{12}^{}$, $\epsilon_{13}^{}$, and
$\epsilon_{23}^{}$ have the same order of magnitude.
This implies that the most stringent constraint on $\epsilon_{ij}^{}$ from $\mu$-$e$
conversion,  \,$|\epsilon_{12}^{}|<1.7\times 10^{-7}$,\,  would translate into
$\epsilon_{13}^{}$ and $\epsilon_{23}^{}$ values that are less than \,$10^{-6}$\,
for all the cases above.  Such constraints, being much smaller than those obtained
from  \,$\tau\to\pi^0 e$\, and \,$\tau\to\mu\bar\mu\mu$,\, would make studies of
FCNC's in $\tau$ decays uninteresting.

Here we would like to point out that it is possible to find solutions for $U_{\nu N}$ that
produce $\epsilon_{12}^{}$ which is very small and, at the same time, $\epsilon_{13,23}^{}$
which are close to current experimental bounds.
We give first an example with large $\epsilon_{13}^{}$ and then another with
large $\epsilon_{23}^{}$.
As in the previous section, to illustrate each of these cases we employ Eq.~(\ref{U0gen})
for the form of $U_0^{}$, choosing the appropriate $V_{11,12,13}$ subject to Eq.~(\ref{MV2})
and adjusting $\kappa$, with the values of $a$, $b$, and $c$ being fixed from experimental data.

In discussing our examples below, we adopt again the tri-bimaximal form
\,$U_{\rm PMNS}=U_{\rm tribi}$.\,
We find as a consequence that the general, symmetric form of the light-neutrino mass matrix
\begin{eqnarray}
m_\nu^{} \,\,=\, \left(\begin{array}{ccc} u \,&\, v \,&\, x \vspace{0.5ex} \\
v & w & y \vspace{0.5ex} \\ x & y & z \end{array}\right)
\end{eqnarray}
can satisfy the diagonalization relation
\,$\hat m_\nu^{}={\rm diag}\bigl(m_{\nu_1},m_{\nu_2},m_{\nu_3}\bigr)=
U_{\rm PMNS}^\dagger m_\nu^{} U_{\rm PMNS}^*$\, only if
\begin{eqnarray}
x \,\,=\,\, v ~, \hspace{5ex} y \,\,=\,\, u+v-w ~, \hspace{5ex} z \,\,=\,\, w ~,
\end{eqnarray}
resulting in the eigen-masses
\begin{eqnarray}
m_{\nu_1}^{} \,\,=\,\, u-v ~, \hspace{5ex} m_{\nu_2}^{} \,\,=\,\, u+2v ~, \hspace{5ex}
m_{\nu_3}^{}  \,\,=\,\, -u-v+2 w ~.
\end{eqnarray}
As can be seen in what follows, this requirement for the elements of $m_\nu^{}$ puts
limitations on the range of choices for the form of $U_{\nu N}$ on the right-hand side of
\,$m_\nu^{}=-U_{\nu N}^{}M_\Sigma^{}U_{\nu N}^{\rm T}$,\, from Eq.~(\ref{UmU}).

\subsection{Solutions with suppressed \boldmath $\epsilon_{12}^{}$ and large $\epsilon_{23}^{}$}

For the case with large $\epsilon_{23}^{}$, we obtain a desired solution with the choice
\,$U_{\nu N}^{}=U_0^e+U_{\alpha\beta\gamma}^e+U_\delta^e$\,  where
\begin{eqnarray} \label{u0e}
U_0^e \,\,=\, \left(\begin{array}{ccc} 0 & 0 & 0 \vspace{0.5ex} \\
0 \,&\, a \,&\, ia \vspace{0.5ex} \\ 0 & b & ib \end{array}\right) \!{\cal R}~, \hspace{5ex}
U_{\alpha\beta\gamma}^e \,\,=\, \left(\begin{array}{ccc} \alpha \,&\, 0 & 0 \vspace{0.5ex} \\
0 & 0 \,&\, 0 \vspace{0.5ex} \\ 0 & 0 & 0 \end{array}\right) \!{\cal R} ~, \hspace{5ex}
U_\delta^e \,\,=\, \left(\begin{array}{ccc} 0 \,&\, \delta_{12}^{} \,&\, 0 \vspace{0.5ex} \\
0 & \delta_{22}^{} & 0 \\ 0 & \delta_{32}^{} & 0 \end{array} \right) \!{\cal R} ~.
\end{eqnarray}
The results are
\begin{eqnarray}
b \,\,=\,\, a ~, \hspace{5ex}
\delta_{22}^{} \,\,=\,\, \delta_{32}^{} \,\,=\,\, \frac{a\,\delta_{12}^{}+\alpha^2}{4 a}~,
\end{eqnarray}
with the eigen-masses given by
\begin{eqnarray}
\hat m_\nu \,\,=\,\, {\rm diag} \bigl( a\,\delta_{12}^{}-\alpha^2,\,
-2 a\,\delta_{12}^{}-\alpha^2,\, 0\bigr) m_N^{} ~,
\end{eqnarray}
and so this is an inverted-hierarchy case with \,$m_{\nu_3}=0$.\,
Numerically, equating the other two eigen-masses to \,$m_{\nu_1}=0.049$\,eV\, and
\,$m_{\nu_2}=0.0498$\,eV,\,  we extract
\begin{eqnarray}
\alpha^2 \,\,=\,\, -4.9\times 10^{-11}~\frac{\rm GeV}{m_N^{}}~, \hspace{5ex}
\delta_{12}^{} \,\,=\,\, \frac{-2.6\times 10^{-13}{\rm\,GeV}}{a\,m_N^{}}~,
\end{eqnarray}
implying
\begin{eqnarray} \label{e1}
\epsilon \,\,=\, \left(\begin{array}{ccc} 0 \,&\, 0 \,&\, 0 \vspace{0.5ex} \\
 0 & 1 & 1 \vspace{0.5ex} \\ 0 & 1 & 1
\end{array}\right) |a|^2\, \bigl(r_2^{} + r_3^{}\bigr) ~.
\end{eqnarray}
Thus, the constraint  \,$|\epsilon_{23}^{}|=|\epsilon_{\mu\tau}^{}|<4.9\times 10^{-4}$\,
from  \,$\tau\to\mu\bar\mu\mu$\, decays translates into
\,$|a|\sqrt{r_2^{} + r_3^{}}<2.2\times 10^{-2}$.\,

\subsection{Solutions with suppressed \boldmath $\epsilon_{12}^{}$ and large $\epsilon_{13}^{}$}

For the example with large $\epsilon_{13}^{}$, we find that choosing
\,$U_{\nu N}^{}=U_0^f+U_{\alpha\beta\gamma}^f+U_\delta^e$,\,  with
\begin{eqnarray} \label{u0f}
U_0^f \,\,=\, \left( \begin{array}{ccc} 0 \,&\, a \,&\, i a \vspace{0.5ex} \\
0 & 0 & 0 \vspace{0.5ex} \\ 0 & b & i b \end{array}\right) \!{\cal R}~, \hspace{5ex}
U_{\alpha\beta\gamma}^f \,\,=\, \left(\begin{array}{ccc} \alpha \,&\, 0 & 0 \vspace{0.5ex} \\
\beta & 0 \,&\, 0 \vspace{0.5ex} \\ 0 & 0 & 0 \end{array}\right) \!{\cal R}~,
\end{eqnarray}
yields the desired results.\footnote{We remark that $U_0^e$ in Eq.~(\ref{u0e}) or $U_0^f$ in
Eq.~(\ref{u0f}) is basically $U_0^d$ in Eq.~(\ref{U0}) with its $a$ or $b$ set to zero, respectively.}
This particular choice allows all the three light-neutrinos to have nonzero masses.
Since the expressions for the parameters are too lengthy to display here, we only show
their numerical values.
Thus, taking  \,$m_{\nu_2}=0.1$\,eV,\,  we have as possible solutions
\begin{eqnarray}
{\rm(i)} \;\; b &=& (0.0013+1.03\,i)a ~, \hspace{5ex}
\alpha^2 \,=\, \frac{(2.8+0.14\,i)\rm\,GeV}{10^{13}~m_N^{}}~, \hspace{5ex}
\beta^2 \,=\, \frac{-1.1\rm~GeV}{10^{10}~m_N^{}}~, \nonumber \\
\delta_{12}^{}  &=&  \frac{-5.0\rm~GeV}{10^{11}~a\,m_N^{}}~, \hspace{5ex}
\delta_{22}^{} \,=\, \frac{(0.01-5.4\,i)\rm\,GeV}{10^{12}~a\,m_N^{}}~, \hspace{5ex}
\delta_{32}^{} \,=\, \frac{(-0.01+ 5.1\,i)\rm\,GeV}{10^{11}~a\,m_N^{}} \hspace*{4ex}
\end{eqnarray}
in the normal-hierarchy case
$\bigl($with \,$m_{\nu_1}=0.0996$\,eV\, and \,$m_{\nu_3}=0.111$\,eV$\bigr)$ and
\begin{eqnarray}
{\rm(ii)} \;\; b &=& (0.0012-0.96\,i)a ~, \hspace{5ex}
\alpha^2 \,=\, \frac{(5.0+0.17\,i)\rm\,GeV}{10^{13}~m_N^{}}~, \hspace{5ex}
\beta^2 \,=\, \frac{-9.3\rm~GeV}{10^{11}~m_N^{}}~, \nonumber \\
\delta_{12}^{}  &=&  \frac{-5.0\rm~GeV}{10^{11}~a\,m_N^{}}~, \hspace{5ex}
\delta_{22}^{} \,=\, \frac{(-0.01-6.8\,i)\rm\,GeV}{10^{12}~a\,m_N^{}}~, \hspace{5ex}
\delta_{32}^{} \,=\, \frac{(-0.01-4.8\,i)\rm\,GeV}{10^{11}~a\,m_N^{}} \hspace*{4ex}
\end{eqnarray}
in the inverted-hierarchy case
$\bigl($with \,$m_{\nu_1}=0.0996$\,eV\, and \,$m_{\nu_3}=0.0867$\,eV$\bigr)$.
These numbers lead to, respectively,
\begin{eqnarray} \label{e2}
{\rm(i)} \;\; \epsilon \,\,=\,
\left(\begin{array}{ccc} 1 &\, 0 \,&\, 0.001-1.0\,i \vspace{0.5ex} \\
0 & 0 & 0 \vspace{0.5ex} \\ 0.001+1.0\,i \,& 0 & 1.1
\end{array}\right) |a|^2\, \bigl(r_2^{} + r_3^{}\bigr) ~,
\end{eqnarray}
\begin{eqnarray} \label{e3}
{\rm(ii)} \;\; \epsilon \,\,=\,
\left(\begin{array}{ccc} 1 &\, 0 \,&\, 0.001+0.96\,i \vspace{0.5ex} \\
0 & 0 & 0 \vspace{0.5ex} \\ 0.001-0.96\,i \,& 0 & 0.93
\end{array}\right) |a|^2\, \bigl(r_2^{} + r_3^{}\bigr) ~.
\end{eqnarray}
The bound  \,$|\epsilon_{13}^{}|=|\epsilon_{e\tau}^{}|<4.2\times 10^{-4}$\, from
\,$\tau\to\pi^0 e$\, decays then implies  \,$|a|\sqrt{r_2^{}+r_3^{}}<2.0\times10^{-2}$\,
in the two cases.

\subsection{Some implications for testing type-III seesaw at the LHC}

We have shown that, with the constraints from FCNC transitions as well as from the tiny neutrino
masses, the elements of $U_{\nu N}$ can still be large enough to be measurable at the LHC.
As mentioned earlier, there are also constraints from the electroweak precision data (EWPD)
obtained by measurements of processes involving neutral currents that conserve lepton flavor.
For type-III seesaw, these EWPD bounds are
\,{\footnotesize$\sum$}$_i^{}\,|(U_{\nu N})_{1i}|^2\le 0.00036$,\,
\,{\footnotesize$\sum$}$_i^{}\,|(U_{\nu N})_{2i}|^2\le 0.00029$,\,  and
\,{\footnotesize$\sum$}$_i^{}\,|(U_{\nu N})_{3i}|^2\le 0.00073$\,~\cite{delAguila:2008cj,ewpd}.
Clearly, these are none other than constraints on the diagonal elements of
\,$\epsilon=U_{\nu N}^{}U_{\nu N}^\dagger$.\,
For the examples above, with $\epsilon$ given in Eqs.~(\ref{e1}), (\ref{e2}), and~(\ref{e3}),
the EWPD bounds translate into \,$|a|\sqrt{r_2^{}+r_3^{}}<1.9\times10^{-2}$,\,
which is comparable to the numbers from FCNC constraints.  Assuming that
\,$r_2^{}\sim r_3^{}={\cal O}(1)$,\, we conclude that \,$|a|\lesssim 0.01$.\,

We can now consider how this result translates into the production of $N$ and $E$ at the LHC.
Entertaining the possibility of large light-heavy mixing, we again focus on the single
production of these heavy leptons.
For $N$, the main channel \,$p p\to l N X$\, arising from  \,$q\bar q'\to W^*\to l N$\,
in this case is the same as that in type-I seesaw, and the cross section is already graphed
in Fig.~\ref{csplot}.  For definiteness, we take  \,$r_2^{}=r_3^{}=1$.\,
Accordingly, with the nonzero elements of $U_{\nu N}$ being of order~0.01, the production
cross-section exceeds 1\,fb for masses up to \,$m_N^{}=115$\,GeV.\,
Furthermore, 100\,fb$^{-1}$ of integrated luminosity can yield about 200 heavy neutrinos
having a~100-GeV mass and at least a few of them with \,$m_N^{}=300$\,GeV.\,

The heavy charged lepton $E$ can also be produced through the mixing via
\,$q\bar q\to(Z^*,h^*)\to l^\pm E^\mp$\, and \,$q\bar q'\to W^*\to\nu E$\,
if the $U_{\nu N}$ elements are sizable.
The Lagrangian containing the relevant interactions is given in Eq.~(\ref{LE}).
With a light charged lepton in the final state, the detection of the $l E$ channel would be
easier, and so we focus on it.  Thus, we calculate the cross-section of \,$pp\to l E X$,\, and
the result is indicated by the dashed curve in Fig.~\ref{csplot}, where the $U_{\nu N}$
element associated with the $ZlE$ coupling has been set to unity in the cross section.
This cross section is seen to be comparable to that of \,$pp\to l N X$.\,
With the nonzero elements of $U_{\nu N}$ being 0.01 in size, the \,$pp\to lEX$\, cross-section
stays bigger than 1\,fb for masses up to \,$m_E^{}=115$\,GeV.\,
With 100\,fb$^{-1}$ of integrated luminosity, the production of more than 200 $E$'s of
100-GeV mass is possible and at least a few of them having \,$m_N^{}=300$\,GeV.\,

We should also mention that, since the heavy leptons in type-III seesaw have gauge
interactions, they can be pair produced at the LHC through \,$q\bar q\to Z^*\to E^+E^-$\,
and  \,$q\bar q'\to W^*\to N E$.\,~\cite{Franceschini:2008pz,delAguila:2008cj}.
It has been shown that the heavy leptons with masses as large as 1\,TeV may be discovered
using these modes~\cite{Franceschini:2008pz,delAguila:2008cj}.
In the case of small light-heavy mixing, the decay vertex of a heavy lepton will be detectably
displaced from its production point, and this can serve as a~distinguishing clue for the small
mixing, as the vertex displacement is unlikely to be observable if the mixing is large.
Also, as pointed out earlier in the type-I case, with small light-heavy mixing in type-III
seesaw there is a correlation between the decay rates of the heavy leptons and
the light-neutrino mixing and masses~\cite{li-he}.
This correlation will be changed if the light-heavy mixing is large.
In that case, the single-production channels, $l N$ and~$l E$, discussed above
can provide complementary information about the nature of their interactions.

\begin{figure}[t]
\includegraphics[width=171mm]{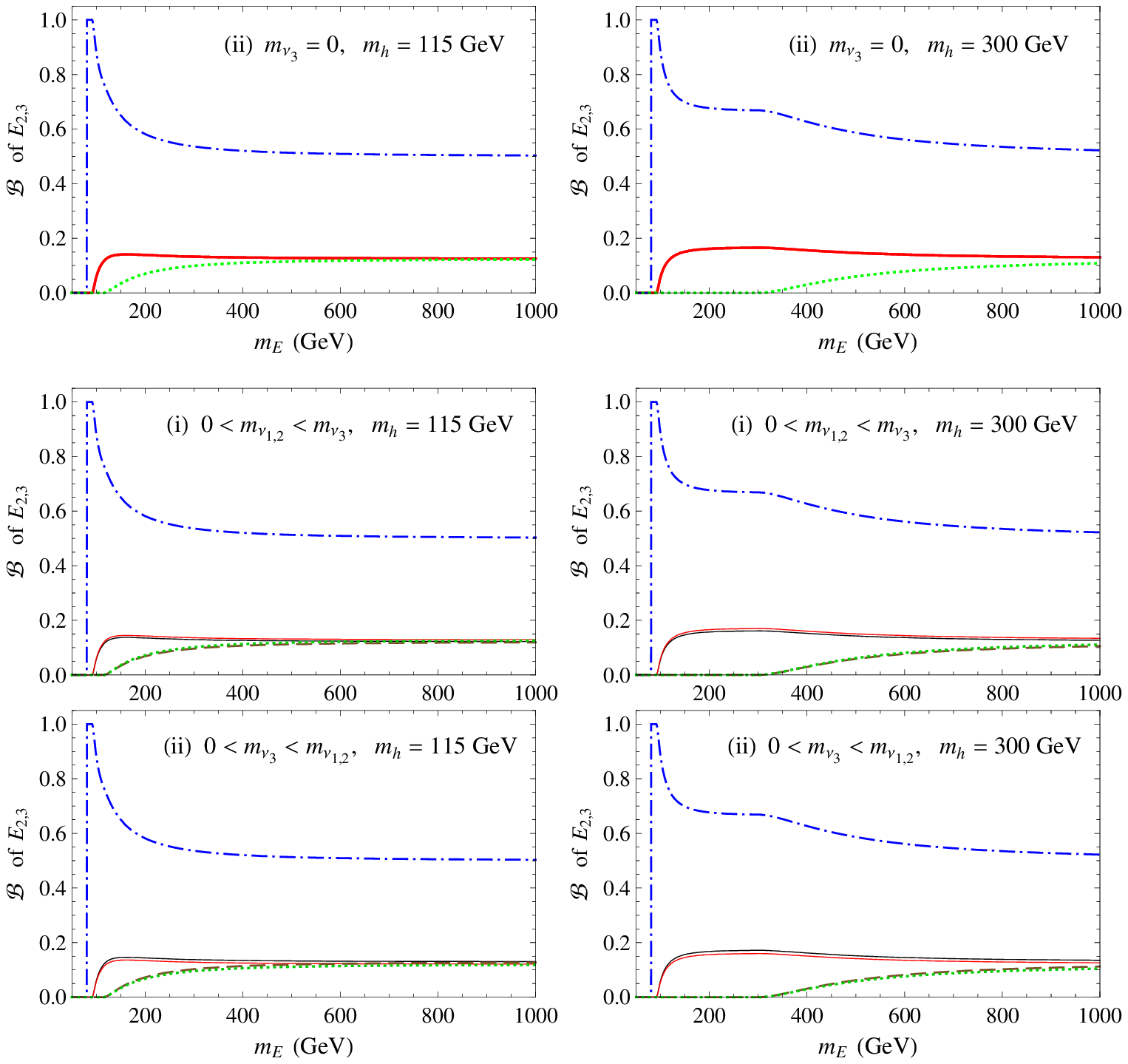}
\caption{Branching ratios $\cal B$ of \,$E\to l Z$ (solid curves), \,$E\to\nu W$
(dot-dashed blue curves), and \,$E\to l h$\, as functions of
the $E$ mass for the sample solutions of $U_{\nu N}$ found in type-III seesaw and Higgs masses
\,$m_h^{}=115$ and 300\,GeV.\,
The top two (bottom four) plots correspond to the examples with one (none) of the light
neutrinos being massless, in either the normal~(i) or inverted~(ii) hierarchy case.
In the top two plots, each red solid curve belongs to both \,$E\to\mu Z$ and $E\to\tau Z$,
which are equal in branching ratio, whereas each green dotted curve belongs to both
\,$E\to\mu h$ and $E\to\tau h$,\, also equal in branching ratio.
In the bottom four plots, each black (red [gray]) solid curve corresponds to \,$E\to e Z$
($E\to\tau Z$) and each brown dashed (green dotted) curve to
\,$E\to e h$~($E\to\tau h$).\label{brplot3}}
\end{figure}

If the LHC does discover the heavy leptons, studying their decay rates and branching ratios
will reveal various information about type-III seesaw, such as the light-heavy mixing,
light-neutrino mixing, and light-neutrino masses.
The relevant formulas for the dominant decay modes can be found in Appendix~\ref{widths}.
We display in Figs.~\ref{brplot2} and~\ref{brplot3} the branching ratios of $N$ and $E$ for
the examples treated in this section.
We note that the curves belonging to the $\nu W$ modes have been obtained after summing over
contributions with $\nu_{1,2,3}^{}$ in the final states, and that certain modes are absent
from the plots because the $U_{\nu N}$ elements associated with them are zero or vanishingly
small.  As in the type-I case, these graphs illustrate that evaluating the branching ratios
could uncover some information on the light-heavy mixing and the light-neutrino mass hierarchy.

\begin{figure}[t]
\includegraphics[width=171mm]{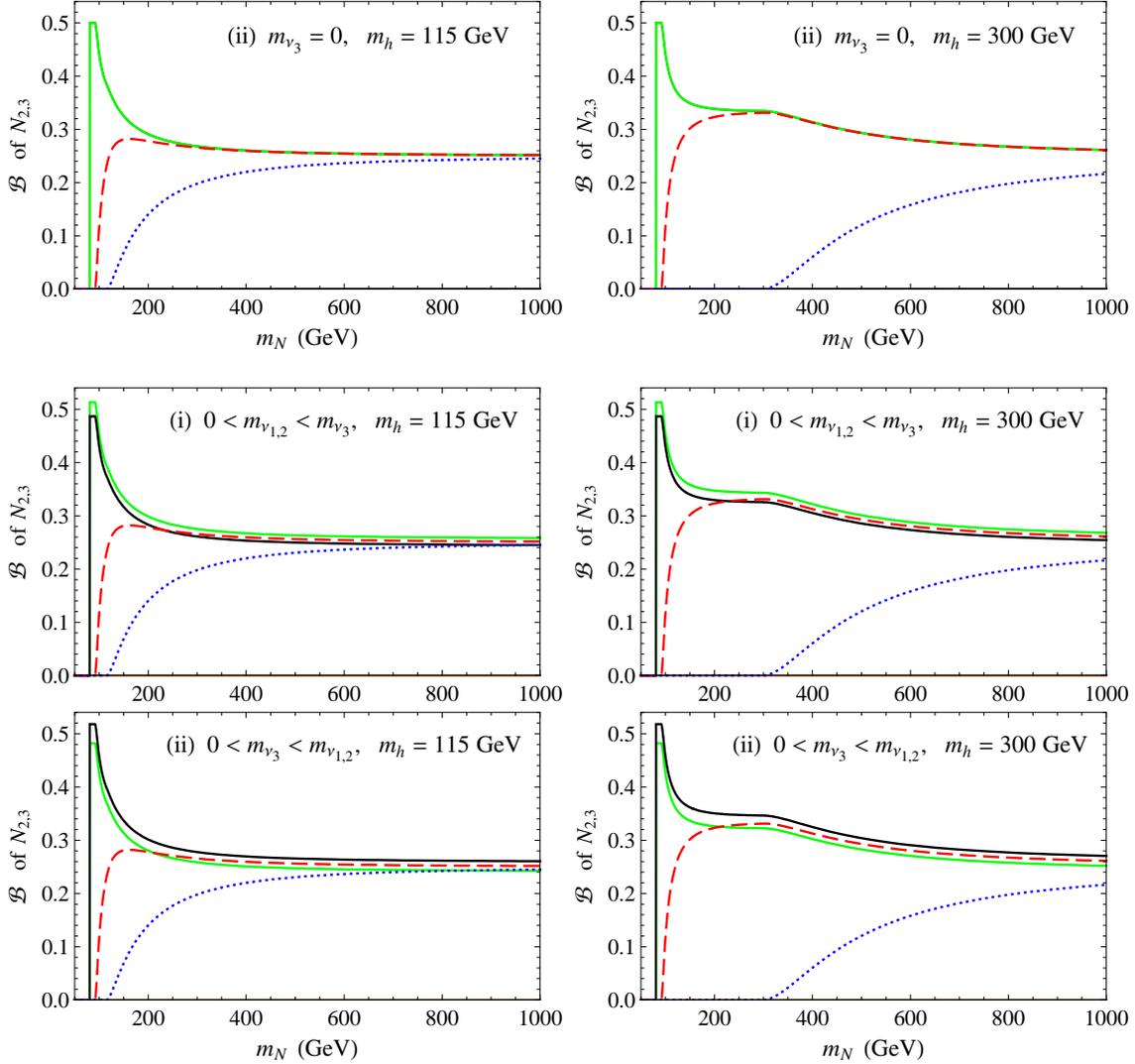}
\caption{Branching ratios $\cal B$ of \,$N\to l W$ (solid curves), \,$N\to\nu Z$
(dashed red curves), and \,$N\to\nu h$ (dotted blue curves)\, as functions of the $N$ mass for
the sample solutions of $U_{\nu N}$ found in type-III seesaw and Higgs masses
\,$m_h^{}=115$ and 300\,GeV.\,
The top two (bottom four) plots correspond to the examples with one (none) of the light
neutrinos being massless, in either the normal~(i) or inverted~(ii) hierarchy case.
In the top two plots, each green solid curve belongs to both \,$N\to\mu W$ and $N\to\tau W$,
which are equal in branching ratio.
In the bottom four plots, each black solid curve refers to \,$N\to e W$\, and each green (gray)
solid curve to only \,$N\to\tau W$.\label{brplot2}} \vspace*{-0.3ex}
\end{figure} 

\section{Conclusions}

In this paper we have explored the possibility of large mixing between the light and heavy
neutrinos in the seesaw scenarios of types I and~III, taking into account constraints from
existing experimental data and considering its implications for testing the models at the LHC.
With just one generation of the neutrinos, it is not possible to have light-heavy
mixing that is sufficiently large to allow the type-I seesaw model to be directly tested at
colliders.  However, with more than one generation of light and heavy neutrinos, the mixing can
be much larger in certain special circumstances, providing more hope for probing the models at
the~LHC. We have investigated this possibility further and presented specific examples in detail.

We have shown that for more than one generation, if the Dirac mass matrix $m_D^{}$ has rank one
and the Majorana mass matrix $M_N$ of the heavy neutrinos has rank three with the appropriate
texture, it is possible to have all the three light neutrinos being massless with
a nontrivial~$m_D^{}$.
The elements of $m_D^{}$ are therefore unconstrained by the light-neutrino masses.
With small corrections added to this special form of $m_D^{}$, it is possible to reproduce
the experimental data on neutrino masses and mixing.

In the type-I seesaw model, we have provided some examples for light-heavy mixing as large as
allowed by current experimental constrains and considered its impact on the production of
the heavy neutrinos at the LHC.
Focusing on the main channel  \,$q\bar q'\to W^*\to l N$,\,  we have found that heavy neutrinos
of masses up to 250\,GeV can be produced by \,$p p\to lNX$\, with a cross section above~1\,fb.
Moreover, with 100\,fb$^{-1}$ of integrated luminosity, the production of over 3000 neutrinos
having 100-GeV mass is possible.

In the type-III seesaw model, the introduction of heavy weak-SU(2)$_{\rm L}$ fermion
triplets makes the phenomenology richer.
New FCNC transitions involving ordinary charged leptons can occur at tree level because of
their interactions with the $Z$ boson.  These FCNC processes lead to strong constraints on
the light-heavy mixing parameterized by $U_{\nu N}$.
If the FCNC parameters calculated from $U_{\nu N}$ are all of the same order of magnitude,
the experimental bound on $\mu$-$e$ conversion renders FCNC decays of the $\tau$ lepton
too small to observe.  This also makes it unlikely for the LHC to detect singly-produced
heavy leptons of the model, $N$ and~$E$.
We have found that it is possible to have solutions for the light-heavy mixing which result
in suppressed $\mu$-$e$ conversion, but large FCNC $\tau$ transitions, and at the same time
allow significant production of the heavy leptons at the LHC.

In type-III seesaw the main channels for single production are \,$q\bar q'\to W^*\to l N$\,
and  \,$q\bar q\to Z^*\to l^\pm E^\mp$,\, corresponding to \,$pp\to l N X$\,  and
\,$pp\to lEX$\,  at the LHC.  We have found in this case that the cross-sections of the two
are comparable.  Thus both $N$ and $E$ can be produced with cross sections higher than~1\,fb
for masses up to 115\,GeV.
With 100\,fb$^{-1}$ of integrated luminosity, the production of more than 200 of these
heavy leptons having a mass of 100\,GeV is possible.
All our estimates above in both seesaw scenarios of types I and~III suggest that there are
interesting prospects for discovering these heavy particles at the LHC and should give
further motivation for carrying out dedicated experimental searches of them.

Last but not least, we have also discussed the significance of studying the decay rates and
branching ratios of the heavy leptons in the seesaw scenarios with large light-heavy mixing.
This will become important once the LHC has seen the heavy leptons.
By examining the rates and branching ratios, one can gain some information about the light-heavy
and/or light-neutrino mixing, as well as the light-neutrino masses.
To help illustrate this, we have presented a number of plots of the dominant branching ratios
of the heavy leptons for the solutions given in our examples.

\acknowledgments \vspace*{-1ex}
This work was supported in part by NSC and NCTS.
We thank K.~Babu, T.~Han, and E.~Ma for discussions and for bringing some of the papers in
Ref.~\cite{large-mix} to our attention.

\appendix

\section{Heavy-lepton partial decay widths\label{widths}} \vspace*{-1ex}

In type-I seesaw, one can extract from Eq.~(\ref{L'}) the couplings of each heavy neutrino $N$
to SM particles.  For the $N$ decay modes expected to be dominant, if kinematically allowed,
one can then derive the partial widths
\begin{eqnarray} \label{NlW}
\Gamma\bigl(N_i^{}\to l_j^+W^-\bigr) \,\,=\,\,
\Gamma\bigl(N_i^{}\to l_j^-W^+\bigr) \,\,=\,\,
\frac{g^2\,\bigl|(U_{\nu N})_{ji}\bigr|^2\,m_N^3}{64\pi\,m_W^2} \Biggl( 1
- \frac{3 m_W^4}{m_N^4} + \frac{2 m_W^6}{m_N^6}\Biggr) ~,
\end{eqnarray} 
\begin{eqnarray}
\Gamma\bigl(N_i^{}\to \nu_j^{}Z\bigr) \,\,=\,\,
\frac{g^2\,\bigl|\bigl(U_{\rm PMNS}^\dagger U_{\nu N}^{}\bigr)_{ji}\bigr|^2\,m_N^3}
     {64\pi\,m_W^2} \Biggl( 1
- \frac{3 m_Z^4}{m_N^4} + \frac{2 m_Z^6}{m_N^6}\Biggr) ~,
\end{eqnarray}
\begin{eqnarray} \label{Nlh}
\Gamma\bigl(N_i^{}\to \nu_j^{}h\bigr) \,\,=\,\,
\frac{g^2\,\bigl|\bigl(U_{\rm PMNS}^\dagger U_{\nu N}^{}\bigr)_{ji}\bigr|^2\,m_N^3}
     {64\pi\, m_W^2} \Biggl( 1 - \frac{m_h^2}{m_N^2}\Biggr)^{\!2} ~,
\end{eqnarray}
having made use of the Majorana nature of $\nu$ and $N$, as well as the relations
\,$m_W^{}=c_{\rm w}^{}m_Z^{}=g v/2$.\,  It follows that, since
\,$U_{\nu N}^\dagger U_{\nu N}^{}\simeq\hat M_N^{-1}m_D^{}m_D^\dagger\hat M_N^{-1}=\epsilon'$,\,
the sum of widths  \,$\Sigma_{j=1,2,3}\Gamma\bigl(N_i^{}\to l_j W\bigr)$\, is independent of
the individual elements of $U_{\nu N}$.
The same is true for  \,$\Sigma_j^{}\Gamma\bigl(N_i\to\nu_j^{}Z\bigr)$\,
and  \,$\Sigma_j^{}\Gamma\bigl(N_i^{}\to\nu_j^{}h\bigr)$.\,
Furthermore, in the large-mixing case under consideration, where $U_{\nu N}^{}$ becomes
decoupled from the light-neutrino masses, the widths are also independent of them.

In type-III seesaw, the couplings of $N$ to SM particles are the same as those in the type-I
case, and therefore the rates of the dominant decay modes are also given by
Eqs.~(\ref{NlW})-(\ref{Nlh}).
For the heavy charged leptons $E$, the corresponding rates are
\begin{eqnarray}
\Gamma\bigl(E_i^+\to\bar\nu_j^{}W^+\bigr) \,\,=\,\,
\frac{g^2\,\bigl|\bigl(U_{\rm PMNS}^\dagger U_{\nu N}^{}\bigr)_{ji}\bigr|^2\,m_E^3}
{32\pi\,m_W^2} \Biggl( 1 - \frac{3 m_W^4}{m_E^4} + \frac{2 m_W^6}{m_E^6}\Biggr) ~,
\end{eqnarray}
\begin{eqnarray}
\Gamma\bigl(E_i^+\to l_j^+Z\bigr) \,\,=\,\,
\frac{g^2\,\bigl|(U_{\nu N})_{ji}\bigr|^2\,m_E^3}
     {64\pi\,m_W^2} \Biggl( 1
- \frac{3 m_Z^4}{m_E^4} + \frac{2 m_Z^6}{m_E^6}\Biggr) ~,
\end{eqnarray} 
\begin{eqnarray} \label{Elh}
\Gamma\bigl(E_i^+\to l_j^+h\bigr) \,\,=\,\,
\frac{g^2\,\bigl|(U_{\nu N})_{ji}\bigr|^2\,m_E^3}
     {64\pi\, m_W^2} \Biggl( 1 - \frac{m_h^2}{m_E^2}\Biggr)^{\!2} ~.
\end{eqnarray}
As with the $N$ rates, the sums \,$\sum_{j} \Gamma\bigl(E_i^{+}\to \bar \nu_j W^+\bigr)$,\,
\,$\sum_{j} \Gamma\bigl(E_i^{+}\to l_j^{+} Z\bigr)$,\, and
\,$\sum_{j} \Gamma\bigl(E_i^{+}\to l_j^{+} h\bigr)$\, are independent of the individual
elements of $U_{\nu N}$.  Our results above for the $N$ and $E$ widths agree with those given
in Ref.~\cite{Franceschini:2008pz,delAguila:2008cj}.

To gain further information on the detectability of $N$ and $E$ at the LHC in the large-mixing
case, we can consider how far they may travel in the center-of-mass (c.m.) frame after being
singly produced and before decaying.
This requires knowing their lifetimes, which can be estimated by employing
Eqs.~(\ref{NlW})-(\ref{Elh}), along with the numerical values of $U_{\nu N}$ elements found
in our examples.
We also need the speeds of $N$ and $E$ in the c.m. frame corresponding to the largest
energy available, which is 7\,TeV, and the smallest $m_{N,E}^{}$ values above their decay
thresholds, as we look for the largest distances that $N$ and $E$ can travel.
Putting together the numbers and avoiding $m_N^{}$ values close to the \,$N\to e W$\, threshold,
which could lead to an arbitrarily small total-width, we find that for \,$m_N^{}>83$\,GeV\,
the distances traveled by $N$ are less than \,$10^{-5}$\,cm\, in either type-I or -III seesaw.
Similarly, excluding $m_E^{}$ values close to the \,$E\to\nu W$\, threshold,
we find that the distances traveled by $E$ are below \,$10^{-4}$\,cm\, for \,$m_E^{}>81$\,GeV.\,
We conclude that, for the examples considered in this paper, $N$ and $E$ are highly likely
to decay well inside the detector.
For comparison, in the case of small light-heavy mixing, $N$ and $E$ in type-III seesaw can
typically travel up to a~few or tens of centimeters~\cite{Franceschini:2008pz}.

\section{Solutions in type-I seesaw with general \boldmath$U_{\rm PMNS}$\label{upmns}}

We provide here, using $U_{\rm PMNS}$ obtained from experimental data, the numerical counterparts
of Eqs.~(\ref{i0}), (\ref{ii0}), (\ref{i0'}), and~(\ref{ii0'}), which were calculated using
$U_{\rm tribi}$ in Eq.~(\ref{Utb}).
For numerical input relevant to $U_{\rm PMNS}$ and the light-eutrino masses, we adopt the
central values of
\,$\Delta m_{21}^2 = \bigl(7.65^{+0.23}_{-0.20}\bigr)\times 10^{-5}$\,eV$^2$, \,
$\bigl|\Delta m_{31}^2\bigr| =\bigl(2.40^{+0.12}_{-0.11}\bigr)\times 10^{-3}$\,eV$^2$, \,
$\sin^2\theta_{12}^{} = 0.304^{+0.022}_{-0.016}$, \,
$\sin^2\theta_{23}^{} = 0.50^{+0.07}_{-0.06}$,\, and
\,$\sin^2 \theta_{13}^{} = 0.010^{+0.016}_{-0.011}$,\,
obtained from a recent fit to global neutrino data~\cite{Schwetz:2008er}.
For the resulting $U_{\rm PMNS}$, we keep the same sign convention as that for $U_{\rm tribi}$.
It is worth remarking that all the elements of $U_{\rm tribi}$ agree with the corresponding ones
of the empirical $U_{\rm PMNS}$ within~1.5$\sigma$.\,

Thus, if one of the light-neutrino masses vanishes, we find as possible solutions for
\,$U_{\nu N}^{}=U_0^a+U_\delta^{}$\,
\begin{eqnarray} \label{i}
{\rm(i)} \;\; a &=& (0.55 + 0.24\,i)\bar b ~, \hspace{5ex}
b \,=\, ( 0.55 + 1.7\, i)\bar b ~, \hspace{5ex}
c \,=\, (-0.63 + 1.7\, i)\bar b ~, \nonumber \\
\delta_{11}^{} &=&
\frac{(-2.4 + 1.0\,i)\,\rm GeV}{10^{12}\,\bar b\, m_N^{}} ~, \hspace{3ex}
\delta_{21}^{} \,=\,
\frac{(-2.4 + 7.3\,i)\,\rm GeV}{10^{12}\,\bar b\, m_N^{}} ~, \hspace{3ex}
\delta_{31}^{} \,=\,
\frac{( 2.8 + 7.3\,i)\,\rm GeV^{\vphantom{\sum}}}{10^{12}\,\bar b\, m_N^{}} \hspace*{5ex}
\end{eqnarray}
in the normal-hierarchy case and \newpage \noindent
\begin{eqnarray} \label{ii}
{\rm(ii)} \;\; a &=& (0.55 + 0.82\,i)\bar b ~, \hspace{5ex}
b \,=\, ( 0.55 - 0.45)\bar b ~, \hspace{5ex}
c \,=\, (-0.63 + 0.33)\bar b ~, \nonumber \\
\delta_{11}^{} &=&
\frac{(-1.4+2.0\,i)\,\rm GeV}{10^{11}\,\bar b\, m_N^{}} ~, \hspace{3ex}
\delta_{21}^{} \,=\,
\frac{(-1.4-1.1\,i)\,\rm GeV}{10^{11}\,\bar b\, m_N^{}} ~, \hspace{3ex}
\delta_{31}^{} \,=\,
\frac{( 1.6+0.8\,i)\,\rm GeV^{\vphantom{\sum}}}{10^{11}\,\bar b\, m_N^{}} \hspace*{5ex}
\end{eqnarray}
in the inverted-hierarchy case, with the other $\delta$'s vanishing.  These numbers are
similar to those in Eqs.~(\ref{i0}) and~(\ref{ii0}) in accord with expectation.

If all the light-neutrino masses are not zero, we get
\begin{eqnarray} \label{i'}
{\rm(i)} \;\; a &=& (0.83 + 0.11\,i)\bar a ~, \hspace{5ex}
b \,=\, (-0.45 + 0.74\,i)\bar a ~, \hspace{5ex}
c \,=\, ( 0.33 + 0.74\,i)\bar a ~, \nonumber \\ \vphantom{\sum_|^|}
\alpha  &=&  \frac{-8.3 - 6.5\,i}{10^6\,\sqrt{m_N^{}/\rm GeV}} ~, \hspace{4ex}
\beta  \,=\, \frac{ 4.5 -  13\,i}{10^6\,\sqrt{m_N^{}/\rm GeV}} ~, \hspace{4ex}
\gamma \,=\, \frac{-3.3 - 1.1\,i}{10^6\,\sqrt{m_N^{}/\rm GeV}} ~, \\
\delta_{12}^{} &=&
\frac{(-8.3 - 5.5\,i)\,\rm GeV}{10^{11}\,\bar a\, m_N^{}} ~, \hspace{4ex}
\delta_{22}^{} \,=\,
\frac{( 4.5 - 5.5\,i)\,\rm GeV}{10^{11}\,\bar a\, m_N^{}} ~, \hspace{4ex}
\delta_{32}^{} \,=\,
\frac{(-3.3 + 6.3\,i)\,\rm GeV}{10^{11}\,\bar a\, m_N^{}} ~, \nonumber
\end{eqnarray}
\begin{eqnarray} \label{ii'}
{\rm(ii)} \;\; a &=& (0.83 + 0.09\,i)\bar a ~, \hspace{5ex}
b \,=\, (-0.45 + 0.66\,i)\bar a ~, \hspace{5ex}
c \,=\, ( 0.33 + 0.66\,i)\bar a ~, \nonumber \\ \vphantom{\sum_|^|}
\alpha  &=&  \frac{-8.3 - 6.4\,i}{10^6\,\sqrt{m_N^{}/\rm GeV}} ~, \hspace{4ex}
\beta  \,=\, \frac{ 4.5 -  12\,i}{10^6\,\sqrt{m_N^{}/\rm GeV}} ~, \hspace{4ex}
\gamma \,=\, \frac{-3.3 - 2.6\,i}{10^6\,\sqrt{m_N^{}/\rm GeV}} ~, \\
\delta_{12}^{} &=&
\frac{(-8.3 - 5.5\,i)\,\rm GeV}{10^{11}\,\bar a\, m_N^{}} ~, \hspace{4ex}
\delta_{22}^{} \,=\,
\frac{( 4.5 - 5.5\,i)\,\rm GeV}{10^{11}\,\bar a\, m_N^{}} ~, \hspace{4ex}
\delta_{32}^{} \,=\,
\frac{(-3.3 + 6.3\,i)\,\rm GeV}{10^{11}\,\bar a\, m_N^{}} \nonumber
\end{eqnarray}
as possible solutions for  \,$U_{\nu N}^{}=U_0^d+U_{\alpha\beta\gamma}^{}+U_\delta^{}$\,
in the (i) normal- and (ii) inverted-hierarchy cases, respectively, with the other $\delta$'s
vanishing.  These numbers are also similar to those in Eqs.~(\ref{i0'}) and~(\ref{ii0'}).


\begin{thebibliography}{99}

\bibitem{pdg}  
  C.~Amsler {\it et al.}  [Particle Data Group],
  Phys.\ Lett.\  B {\bf 667}, 1 (2008)
and 2009 partial update for the 2010 edition.

\bibitem{zee} A.~Zee,
  Phys.\ Lett.\  B {\bf 93}, 389 (1980)
  [Erratum-ibid.\  B {\bf 95}, 461 (1980)];
  K.S.~Babu,
  Phys.\ Lett.\  B {\bf 203}, 132 (1988);
  E.~Ma,
  Phys.\ Rev.\ Lett.\  {\bf 81}, 1171 (1998)
  [arXiv:hep-ph/9805219].
  A.~Pilaftsis,    
  Z.\ Phys.\  C {\bf 55}, 275 (1992)
  [arXiv:hep-ph/9901206].

\bibitem{seesaw1}
  P.~Minkowski,
  Phys.\ Lett.\  B {\bf 67}, 421 (1977);
T.~Yanagida, in {\it Proceedings of the Workshop on the Unified Theory and the Baryon Number in
the Universe}, edited by O.~Sawada and A.~Sugamoto (KEK, Tsukuba, 1979), p.~95;
M.~Gell-Mann, P.~Ramond, and R.~Slansky, in {\it Supergravity},
edited by P.~van Nieuwenhuizen and D.~Freedman (North-Holland, Amsterdam, 1979), p.~315;
S.L.~Glashow, in {\it Proceedings of the 1979 Cargese Summer Institute on Quarks and Leptons},
edited by M.~Levy {\it et al}. (Plenum Press, New York, 1980), p. 687;
  R.N.~Mohapatra and G.~Senjanovic,
  Phys.\ Rev.\ Lett.\  {\bf 44}, 912 (1980).

\bibitem{seesaw2}
W.~Konetschny and W.~Kummer,
  Phys.\ Lett.\  B {\bf 70}, 433 (1977);
%
 T.P.~Cheng and L.F.~Li,
  Phys.\ Rev.\  D {\bf 22}, 2860 (1980);
%
 G.~Lazarides, Q.~Shafi, and C.~Wetterich,
  Nucl.\ Phys.\  B {\bf 181}, 287 (1981);
%
 J.~Schechter and J.W.F.~Valle,
  Phys.\ Rev.\  D {\bf 22}, 2227 (1980);
%
 R.N.~Mohapatra and G.~Senjanovic,
  Phys.\ Rev.\  D {\bf 23}, 165 (1981).

\bibitem{Foot:1988aq}
  R.~Foot, H.~Lew, X.G.~He, and G.C.~Joshi,
  Z.\ Phys.\  C {\bf 44}, 441 (1989).

\bibitem{colliders} 
  T.~Han and B.~Zhang,
  Phys.\ Rev.\ Lett.\  {\bf 97}, 171804 (2006)
  [arXiv:hep-ph/0604064];
  B.~Bajc, M.~Nemevsek, and G.~Senjanovic,
  Phys.\ Rev.\  D {\bf 76}, 055011 (2007)
  [arXiv:hep-ph/0703080];
  W.~Chao, Z.G.~Si, Z.Z.~Xing, and S.~Zhou,   
  Phys.\ Lett.\  B {\bf 666}, 451 (2008)
  [arXiv:0804.1265 [hep-ph]];
  P.~Fileviez Perez, T.~Han, G.Y.~Huang, T.~Li, and K.~Wang,
  Phys.\ Rev.\  D {\bf 78}, 015018 (2008)
  [arXiv:0805.3536 [hep-ph]];
  F.~del Aguila and J.A.~Aguilar-Saavedra,
  Phys.\ Lett.\  B {\bf 672}, 158 (2009)
  [arXiv:0809.2096 [hep-ph]];
  Z.Z.~Xing,
  Int.\ J.\ Mod.\ Phys.\  A {\bf 24}, 3286 (2009)
  [arXiv:0901.0209 [hep-ph]];
  A.~Atre, T.~Han, S.~Pascoli, and B.~Zhang,
  JHEP {\bf 0905}, 030 (2009)
  [arXiv:0901.3589 [hep-ph]];
  N.~Haba, S.~Matsumoto, and K.~Yoshioka,
  Phys.\ Lett.\  B {\bf 677} (2009) 291
  [arXiv:0901.4596 [hep-ph]];
  A.~Arhrib, B.~Bajc, D.K.~Ghosh, T.~Han, G.Y.~Huang, I.~Puljak, and G.~Senjanovic,
  arXiv:0904.2390 [hep-ph];
  P.~Bandyopadhyay, S.~Choubey, and M.~Mitra,
  arXiv:0906.5330 [hep-ph];
  W.~Chao, Z.G.~Si, Y.J.~Zheng, and S.~Zhou,    
  arXiv:0907.0935 [hep-ph].

\bibitem{del Aguila:2007em}
  F.~del Aguila, J.~A.~Aguilar-Saavedra and R.~Pittau,
  JHEP {\bf 0710}, 047 (2007)
  [arXiv:hep-ph/0703261].

\bibitem{Franceschini:2008pz}
  R.~Franceschini, T.~Hambye, and A.~Strumia,
  Phys.\ Rev.\  D {\bf 78}, 033002 (2008)
  [arXiv:0805.1613 [hep-ph]].

\bibitem{delAguila:2008cj}
  F.~del Aguila and J.A.~Aguilar-Saavedra,
  Nucl.\ Phys.\  B {\bf 813}, 22 (2009)
  [arXiv:0808.2468 [hep-ph]].

\bibitem{large-mix}
  W.~Buchmuller and D.~Wyler,
  Phys.\ Lett.\  B {\bf 249}, 458 (1990);
  W.~Buchmuller and C.~Greub,
  Nucl.\ Phys.\  B {\bf 363}, 345 (1991);
  J.~Gluza,
  Acta Phys.\ Polon.\  B {\bf 33}, 1735 (2002)
  [arXiv:hep-ph/0201002];
  G.~Ingelman and J.~Rathsman,
  Z.\ Phys.\  C {\bf 60}, 243 (1993);
  A.~Pilaftsis,
  Phys.\ Rev.\ Lett.\  {\bf 95}, 081602 (2005)
  [arXiv:hep-ph/0408103];
  A.~Pilaftsis and T.E.J.~Underwood,
  Phys.\ Rev.\  D {\bf 72}, 113001 (2005)
  [arXiv:hep-ph/0506107];
J.~Kersten and A.Y.~Smirnov,
  Phys.\ Rev.\  D {\bf 76}, 073005 (2007)
  [arXiv:0705.3221 [hep-ph]];
E.~Ma,
  Mod.\ Phys.\ Lett.\  A {\bf 24}, 2161 (2009)
  [arXiv:0904.1580 [hep-ph]];
Z.Z.~Xing,
  Prog.\ Theor.\ Phys.\ Suppl.\  {\bf 180}, 112 (2010)
  [arXiv:0905.3903 [hep-ph]].

\bibitem{pmns}
  Z.~Maki, M.~Nakagawa, and S.~Sakata,
  Prog.\ Theor.\ Phys.\  {\bf 28}, 870 (1962);
  B.~Pontecorvo,
  Sov.\ Phys.\ JETP {\bf 26} (1968) 984
  [Zh.\ Eksp.\ Teor.\ Fiz.\  {\bf 53} (1967) 1717].

\bibitem{He:2009xd}
  X.G.~He and E.~Ma,
  Phys.\ Lett.\  B {\bf 683}, 178 (2010)
  [arXiv:0907.2737 [hep-ph]].

\bibitem{Schwetz:2008er}
  T.~Schwetz, M.~Tortola, and J.W.F.~Valle,
  New J.\ Phys.\  {\bf 10}, 113011 (2008)
  [arXiv:0808.2016 [hep-ph]];
  M.~Maltoni and T.~Schwetz,
  arXiv:0812.3161 [hep-ph].
Other recent fits are given by
  M.C.~Gonzalez-Garcia and M.~Maltoni,
  Phys.\ Rept.\  {\bf 460}, 1 (2008)
  [arXiv:0704.1800 [hep-ph]];
  G.L.~Fogli {\it et al.},
  Phys.\ Rev.\  D {\bf 78}, 033010 (2008)
  [arXiv:0805.2517 [hep-ph]].

\bibitem{Harrison:2002er}
  P.F.~Harrison, D.H.~Perkins, and W.G.~Scott,
  Phys.\ Lett.\  B {\bf 530}, 167 (2002)
  [arXiv:hep-ph/0202074];
  Z.Z.~Xing,
  Phys.\ Lett.\  B {\bf 533}, 85 (2002)
  [arXiv:hep-ph/0204049];
  X.G.~He and A.~Zee,
  Phys.\ Lett.\  B {\bf 560}, 87 (2003)
  [arXiv:hep-ph/0301092].

\bibitem{ewpd}
  F.~del Aguila, J.~de Blas, and M.~Perez-Victoria,
  Phys.\ Rev.\  D {\bf 78}, 013010 (2008)
  [arXiv:0803.4008 [hep-ph]];
  F.~del Aguila, J.A.~Aguilar-Saavedra, J.~de Blas, and M.~Perez-Victoria,
  arXiv:0806.1023 [hep-ph].

\bibitem{typeI_fcnc}
  S.~Antusch, C.~Biggio, E.~Fernandez-Martinez, M.~B.~Gavela and J.~Lopez-Pavon,
  JHEP {\bf 0610}, 084 (2006)
  [arXiv:hep-ph/0607020].

\bibitem{mstw}
  A.~D.~Martin, W.~J.~Stirling, R.~S.~Thorne and G.~Watt,
  Eur.\ Phys.\ J.\  C {\bf 63}, 189 (2009)
  [arXiv:0901.0002 [hep-ph]];
http://projects.hepforge.org/mstwpdf.

\bibitem{Abada:2007ux}
A.~Abada, C.~Biggio, F.~Bonnet, M.B.~Gavela, and T.~Hambye,
  JHEP {\bf 0712}, 061 (2007)
  [arXiv:0707.4058 [hep-ph]];

\bibitem{Abada:2008ea}
  A.~Abada, C.~Biggio, F.~Bonnet, M.B.~Gavela, and T.~Hambye,
  Phys.\ Rev.\  D {\bf 78}, 033007 (2008)
  [arXiv:0803.0481 [hep-ph]].

\bibitem{He:2009tf}
  X.G.~He and S.~Oh,
  JHEP {\bf 0909}, 027 (2009)
  [arXiv:0902.4082 [hep-ph]];

\bibitem{Arhrib:2009xf}
  A.~Arhrib, R.~Benbrik, and C.H.~Chen,
  arXiv:0903.1553 [hep-ph].

\bibitem{li-he} 
T.~Li and X.G.~He, 
  Phys.\ Rev.\  D {\bf 80}, 093003 (2009)
  [arXiv:0907.4193 [hep-ph]].

\end{thebibliography}
\end{document}